\documentclass[useAMS,usenatbib]{mn2e}
\usepackage{amsmath}
\usepackage{graphicx}
\usepackage{amssymb}

\def\dalemb#1#2{{\vbox{\hrule height.#2pt
        \hbox{\vrule width.#2pt height#1pt \kern#1pt \vrule width.#2pt}
        \hrule height.#2pt}}}

\def\ba{\begin{eqnarray}}
\def\ea{\end{eqnarray}}
\def\be{\begin{equation}}
\def\ee{\end{equation}}

\def\Var{{\rm Var}}
\def\gtorder{\mathrel{\raise.3ex\hbox{$>$}\mkern-14mu
             \lower0.6ex\hbox{$\sim$}}}
\def\ltorder{\mathrel{\raise.3ex\hbox{$<$}\mkern-14mu
             \lower0.6ex\hbox{$\sim$}}}

\def\bb{{\mathfrak b}}
\newcommand{\ellb }{\boldsymbol{\ell }}

\title[Detecting bispectral acoustic oscillations
from inflation using a new flexible estimator]%
{Detecting bispectral acoustic oscillations
from inflation using a new flexible estimator}
\author[Martin Bucher, Bartjan van Tent, and Carla Sofia Carvalho]{
Martin Bucher 
$^{1,2}$\thanks{E-mail: bucher@apc.univ-paris7.fr},
Bartjan van Tent
$^{1}$\thanks{E-mail: vantent@th.u-psud.fr},
and Carla Sofia Carvalho 
$^{1,3}$\thanks{E-mail: carvalho@th.u-psud.fr}\\
$^{1}$Laboratoire de Physique Th\'eorique, 
Universit\'e Paris-Sud 11 et CNRS, 
B\^atiment 210, 91405 Orsay Cedex, France\\
$^{2}$Laboratoire Astroparticules Cosmologie, 
Universit\'e Paris Diderot 7, 
10 rue Alice Domon et L\'eonie Duquet,
75013 Paris, France\\
$^{3}$University of KwaZulu-Natal, Durban, 4041, South Africa}

\begin{document}


\date{LPT-ORSAY 09-94}


\maketitle

\label{firstpage}

\begin{abstract}
We present a new flexible estimator for comparing theoretical
templates for the predicted bispectrum of the CMB anisotropy
to observations. This estimator, based on binning
in harmonic space, generalizes the ``optimal'' estimator 
of Komatsu, Spergel, and Wandelt by allowing an adjustable
weighting scheme for masking possible foreground and other 
contaminants and detecting particular noteworthy features
in the bispectrum. The utility of this estimator is illustrated
by demonstrating how acoustic oscillations in the bispectrum
and other details of the bispectral shape
could be detected in the future PLANCK data provided that $f_{NL}$
is sufficiently large. The character and 
statistical weight of the acoustic
oscillations and the decay tail are described in detail.

\end{abstract}

\begin{keywords}
cosmic microwave background -- early universe -- methods: statistical.
\end{keywords}

\section{Introduction}

At present the primordial blackbody component of the cosmic microwave
background (CMB) appears very nearly Gaussian, as one would expect
from inflationary models for the very early universe. To a first 
approximation, because of the weak coupling of the physics involved,
inflation predicts a nearly scale-invariant spectrum of primordial
density perturbations whose imprint on the CMB is completely characterized
by the power spectrum $C_\ell ^{AB}$, where $\ell $ is the multipole
number and $(A,B=T,E)$, according to the Gaussian distribution:
\ba
P(\{ a_{\ell , m}\} ) & = &
\prod _{\ell , m} (2\pi )^{-1/2} 
\det {}^{-1/2} 
\left[ C_\ell \right] \cr 
&& \times
\exp \left[ -\frac{1}{2}a_{\ell , m}^TC_\ell ^{-1/2}a_{\ell , m}\right]
\ea 
where
\be
a_{\ell , m}=
\begin{pmatrix} 
a_{\ell , m}^T\cr
a_{\ell , m}^E\cr
\end{pmatrix}, 
\qquad
C_\ell =
\begin{pmatrix} 
C_\ell ^{TT}  & C_\ell ^{TE}\\
C_\ell ^{TE}  & C_\ell ^{EE}\\
\end{pmatrix}
\ee
with $T$ denoting temperature and $E$ the $E$-mode polarization.

It has however been noted that 
even in single scalar field inflationary models 
small nonlinear corrections do arise, at lowest 
order in the bispectrum \citep{mald, acquaviva}. 
In models with several scalar fields
even larger and potentially detectable degrees of 
non-Gaussianity are possible 
\citep{ng_from_inflation, RSvT1, RSvT2, narsas, byrnes1, byrnes2, huang}. 

Non-Gaussianity manifests itself in odd $n$-point 
correlation functions or in the connected 
even $n$-point correlation 
functions, from which the trivial part expressible
as combinations of two-point correlation functions
has been subtracted away.  
The extent of departures from Gaussianity can be 
characterized by ratios of higher-order
correlation functions and the appropriate 
combination of two-point correlation functions \citep{bernardeau}. 
The evolution, for the most part linear, of 
the primordial fluctuations of the inflaton field, 
involving both gravity and hydrodynamics,
leads to CMB anisotropies whose statistical properties 
mirror those of the primordial fluctuations. Consequently, by 
studying higher-order correlation functions of the 
CMB anisotropies, we can detect and characterize
any primordial non-Gaussianity.

The lowest order such statistic is the bispectrum, or 
three-point correlation function in Fourier space. 
The bispectrum has been shown to be an optimal statistic 
for measuring non-Gaussianity in the sense that
the signal-to-noise squared of the non-Gaussianity estimator 
based on the three-point correlation function dominates over all 
higher-order estimators \citep{babich}. 
Consequently it would also be significantly easier to constrain. 
Assuming that the characteristic amplitude of the bispectrum is
much larger than that of the trispectrum, or the four-point 
correlation function in Fourier space,
the non-Gaussianity of the density fluctuations from inflation for the 
single-field case
may be approximated by the following ``local'' ansatz for the
gravitational potential \citep{fnl_expansion,acoustic}
\be
\Phi ({\bf x})
=
\Phi _{G}({\bf x})
+f_{NL} \left [
{\Phi _{G}}^2({\bf x})
- \langle {\Phi _{G}}^2({\bf x}) \rangle \right ],
\label{ansatz}
\ee
where the nonlinearity parameter $f_{NL}$ may be considered small.
While the three-point correlator 
of the linear or Gaussian part $\Phi_G$ is zero (because in the quantum
picture it involves the expectation value of a product of three creation and 
annihilation operators), this is no longer true when the quadratic correction
is taken into account. 
The bispectrum will be proportional to $f_{NL},$ which determines 
the size of non-Gaussianity.
While this form is only approximate for the single-field
case, for the observationally more promising multi-field
models for which potentially observable values of $f_{NL}$
can be generated, this ansatz is very accurate because
the bulk of the non-Gaussianity is imprinted after horizon
crossing.

The primordial bispectrum is characterized by an overall amplitude 
$f_{NL}$ and the shape. 
Translational and rotational invariance reduce the bispectrum to a 
function of the lengths only of the three wave vectors. 
A classification of bispectra shapes has been proposed by
\cite{shapeNGcamb}.
Local and warm models peak on squeezed triangles, the equilateral models
on equilateral triangles, and the folded/flat models on 
flattened triangles. 
Models in which perturbations are generated outside the horizon 
produce a bispectrum 
that peaks on squeezed triangle configurations, examples of 
which are multi-field inflation \citep{curvaton,curvaton_b,RSvT2}. 
[For a review, see \cite{bartolo_et_al_review}.]
The bispectrum of standard single-field inflation can be viewed as a 
superposition of the local shape and the equilateral shape, both terms, 
however, being slow-roll suppressed \citep{mald, shapeNGcamb, current_limits}. 
In particular, for the squeezed triangle configurations in single-field 
inflation the non-Gaussian signal would be 
proportional to the tilt of the power spectrum and thus imply a strong 
deviation from scale invariance \citep{theorem,theorem_b}.
A bispectrum peaking on equilateral triangles is predicted in models with
non-standard kinetic terms in the inflaton Lagrangian, which is the case 
of multi-field inflation models such as DBI inflation \citep{dbi}, 
ghost inflation \citep{ghost} or trapped inflation \citep{trapped}.
Flattened-triangle configurations arise from initial conditions 
which deviate from the Bunch-Davies vacuum \citep{folded}.
The limits for $f_{NL}$ of the local shape from the WMAP5
analysis are $-9 < f_{NL}^{local} < 111$ at 95\% confidence (i.e.\ $2\sigma$)
\citep{wmap5}.

After horizon crossing, 
non-linearities in both the gravitational and hydrodynamical 
evolution of the baryon-photon fluid prior to recombination 
\citep{recombination_b,recombination_c,recombination_d,recombination_e},
as well as higher orders in the gravitational potential during 
recombination \citep{recombination}, can generate non-Gaussianity.
Other sources of non-primordial non-Gaussianity include 
secondary anisotropies such as 
weak lensing via the cross-correlation with the unlensed CMB 
arising from the integrated Sachs-Wolf effect \citep{smithzaldar, 
lensingSW, MangilliVerde} 
or the Sunyaev-Zel'dovich effect \citep{lensingZS}, as well as 
foregrounds such as dust, galactic synchrotron radiation and 
unresolved point sources.
Finally there are also instrumental effects, see e.g.\ \cite{donzelli}.
These effects contribute spurious non-Gaussian signals, thus 
biasing the measurement of the primordial signal. It is therefore
important to develop tools to isolate the primordial signal 
from the contaminants.
 
The separation of the CMB and foreground components exploits
their differing spectral and spatial distribution \citep{wmap5_gold}. 
It has been shown that foreground sources distributed in an anisotropic 
manner can mimic a non-Gaussian signal which could be taken for primordial
\citep{carvalho}.
It has been suggested that
the claimed detection of $f_{NL}$ \citep{yadav} might be  
due to residual foregrounds or instrumental contamination 
\citep{pietrobon, cabella, smith}.
Consequently it is important to develop estimators based on the bispectrum
also able to mask out non-primordial parasite non-Gaussianities.

In this paper we shall focus primarily on 
bispectral non-Gaussianity of the ``local'' type just 
defined, although most of the discussion generalizes
straightforwardly to other templates. 
See also \cite{JamesPaul, shapeNGcamb}, whose authors 
are working on another 
bispectral estimator program, specifically aimed at dealing with theoretical 
primordial bispectra of general momentum dependence. 
In fact, their expansion of a general primordial bispectrum in terms of
separable eigenfunctions is one way to easily extend our work beyond the
``local'' type of primordial non-Gaussianity.
Unlike much of the 
literature, where the emphasis is on developing 
an ``optimal'' estimator, where all the data are reduced
to a single number \citep{heavens_a,koma,yadavetal,creminelli}, 
the emphasis in our paper is placed on developing a 
more flexible approach where the data can be divided
in many ways and under which the effects of parasite
non-Gaussianities can be masked in a well-defined way.  
See also \cite{mexican_hat,needlets,needlets_b, 
hikage,munshia,munshib,santos,smith} for other estimators, many not based 
on the bispectrum but on wavelets, needlets, or
Minkowski functionals.

The paper is organized as follows. In section 2 we establish some notation
and discuss some general properties of estimators. A simplified and
intuitive discussion of the nature and distribution of statistical
weight of the "local" bispectral signal is given and it is shown
how to modify templates in order to best mask parasite contaminants.
Section 3 presents a quantitative discussion of the shape of the
"local" signal and in particular the acoustic oscillations are
examined quantitatively. In section 4 the signal-to-noise for
acoustic oscillations in the vertex region (where $\ell _1\ll
\ell _2, \ell _2$) is examined. Section 5 gives the computational
details of a binned bispectral estimator that can be used for
a general template and presents tests validating this estimator.
Finally in section 6 we present some concluding remarks.

\section{Basic Formalism}

\subsection{Reduced bispectrum}

The temperature fluctuations in the CMB $T(\Omega) = \Delta T(\Omega)/T_0$
can be decomposed into spherical harmonics according to 
\be
T(\Omega) = \sum_{\ell m} a_{\ell m} Y_{\ell m}(\Omega)
\ee
with the harmonic coefficients $a_{\ell m}$ given by 
\be
a_{\ell m} = \int d\Omega \, T(\Omega) Y_{\ell m}^*(\Omega).
\ee
The bispectrum on the celestial sphere consists of cubic combinations
of the spherical harmonic coefficients of the form
\be 
b_{\ell_1 \ell_2 \ell_3}^{m_1 m_2 m_3}
= a_{\ell_1 m_1} a_{\ell_2 m_2} a_{\ell_3 m_3}
\ee
whose expectation values may be calculated for a given theory.
However under the assumption of statistical isotropy, these
expectation values are not independent and can be reduced to
quantities depending only on 
$\ell _1,$ $\ell _2,$ and $\ell _3.$
In the sequel, we shall not always
explicitly indicate expectation values, because in many
cases the equations can be interpreted either with 
expectation values or as relating statistics of 
a particular realization.

We may define a manifestly rotationally-invariant
reduced bispectrum in terms of integrals of triple 
products of maximally filtered maps so that
\be
b_{\ell _1 \ell _2 \ell _3} =
\int d\hat \Omega ~
T_{\ell _1}(\hat \Omega ) ~
T_{\ell _2}(\hat \Omega ) ~
T_{\ell _3}(\hat \Omega ) ,
\label{BispecFromMap}
\ee
where the maximally filtered map is defined as
\be 
T_\ell (\hat \Omega )=\sum _{m=-\ell }^{+\ell }
a_{\ell m} Y_{\ell m}(\hat \Omega ).
\label{TempMap}
\ee
Using the expression for the Gaunt integral 
\ba
{\cal G}_{\ell_1 \ell_2 \ell_3}^{m_1 m_2 m_3} & = & \int d\Omega ~
Y_{\ell _1m_1} (\Omega )~ Y_{\ell _2m_2} (\Omega )~ Y_{\ell _3m_3} (\Omega )\\
& = & 
\sqrt{ \frac{ (2\ell _1+1) (2\ell _2+1) (2\ell _3+1) }{ 4\pi } } 
\nonumber\\
&& \times
\begin{pmatrix}
\ell _1 &\ell _2 &\ell _3\cr
0&0 &0\cr
\end{pmatrix}
\begin{pmatrix}
\ell _1 &\ell _2 &\ell _3\cr
m_1 &m_2 &m_3\cr
\end{pmatrix},
\nonumber
\ea
we obtain
\ba
b_{\ell _1 \ell _2 \ell _3} & = &
\sqrt{ \frac{ (2\ell _1+1) (2\ell _2+1) (2\ell _3+1) }{ 4\pi } } 
\begin{pmatrix}
\ell _1 &\ell _2 &\ell _3\cr
0&0 &0\cr
\end{pmatrix}\cr
&&\times
\sum _{m_1,m_2,m_3}
\begin{pmatrix}
\ell _1 &\ell _2 &\ell _3\cr
m_1 &m_2 &m_3\cr
\end{pmatrix}
b_{\ell_1 \ell_2 \ell_3}^{m_1 m_2 m_3}.
\ea
As a consequence of the Wigner-Eckart theorem,
$b_{\ell_1 \ell_2 \ell_3}^{m_1 m_2 m_3}$
is proportional to
$
\begin{pmatrix}
\ell _1 &\ell _2 &\ell _3\cr
m_1 &m_2 &m_3\cr
\end{pmatrix}.
$
Using this fact combined with the
$3j$-symbol identity
\be
\sum _{m_1m_2m_3}
\begin{pmatrix}
\ell _1 &\ell _2 &\ell _3\cr
m_1 &m_2 &m_3\cr
\end{pmatrix}
\begin{pmatrix}
\ell _1 &\ell _2 &\ell _3\cr
m_1 &m_2 &m_3\cr
\end{pmatrix}
=1,
\ee
which holds whenever $\ell _1,$ $\ell _2,$ $\ell _3$
satisfy the triangle inequality and the 
parity condition 
$(-1)^{\ell _1+\ell _2+\ell _3}=(+1)$,  
we find that
\ba 
b_{\ell_1 \ell_2 \ell_3}^{m_1 m_2 m_3}
& = &
\sqrt{ 
\frac
{4\pi }
{(2\ell _1+1)(2\ell _2+1)(2\ell _3+1)}
}\\
&&\times 
\begin{pmatrix}
\ell _1 &\ell _2 &\ell _3\cr
0 &0 &0\cr
\end{pmatrix} ^{-1}
\begin{pmatrix}
\ell _1 &\ell _2 &\ell _3\cr
m_1 &m_2 &m_3\cr
\end{pmatrix}
b_{\ell _1 \ell _2 \ell _3 }.
\nonumber
\ea 

Appealing to the flat sky approximation, we may interpret 
(\ref{BispecFromMap})
as the expectation
value for a single triangle with sides of dimension 
$\ell _1,$ $\ell _2,$ and $\ell _3$ multiplied by the number of 
possible such triangles
on the celestial sphere $N_\triangle (\ell _1, \ell _2, \ell _3 ).$ The number
of triangles $N_\triangle (\ell _1, \ell _2, \ell _3 )$ may be calculated 
by considering the variance of 
$b_{\ell _1, \ell _2, \ell _3}$ in a Gaussian theory, which should be given by
\be 
{\rm Var}[b_{\ell _1 \ell _2 \ell _3}]
=N_\triangle (\ell _1, \ell _2, \ell _3 )~
c_{\ell _1} c_{\ell _2} c_{\ell _3}~ g ,
\label{VarExp}
\ee
which may be regarded as the formal definition of
$N_\triangle (\ell _1, \ell _2, \ell _3 ).$
Here $g$ is a combinatorial factor, 
equal to 1, 2, or 6 depending on whether one, two, or three
of the $\ell $ are equal, respectively. 
Using the above formulae, we obtain
\be
N_\triangle (\ell _1, \ell _2, \ell _3)=
\frac{(2\ell_1+1)(2\ell_2+1)(2\ell_3+1)}{4\pi }
\begin{pmatrix}
\ell _1 &\ell _2 &\ell _3\cr
0&0 &0\cr
\end{pmatrix}^2.
\label{NTriExp}
\ee

We now obtain a second useful reduced bispectral quantity, which is
often called the reduced bispectrum in the literature, 
\ba
\bb _{\ell_1 \ell_2 \ell_3} & = &
\frac{b_{\ell_1 \ell_2 \ell_3}}{N_\triangle(\ell _1, \ell _2, \ell _3)}\\
& = &
\sqrt{
\frac%
{4\pi }
{(2\ell_1+1)(2\ell_2+1)(2\ell_3+1)}
}
\begin{pmatrix}
\ell _1 &\ell _2 &\ell _3\cr
0&0 &0\cr
\end{pmatrix}^{-1} \nonumber\\ 
&&\times \sum_{m_1,m_2,m_3}
\left( \begin{array}{ccc}
\ell_1 & \ell_2 & \ell_3\\
m_1 & m_2 & m_3
\end{array}\right)
b_{\ell_1 \ell_2 \ell_3}^{m_1 m_2 m_3}.
\nonumber
\ea
This definition 
is the direct analogue of $b(
{\ellb _1}, 
{\ellb _2}, 
{\ellb _3} 
)$
in the flat-sky approximation.

\subsection{Natural inner product and ideal estimator}

Under the assumptions of
full sky coverage, isotropic instrument noise
characterized by a 
power spectrum $n_\ell$ (where the beam profile 
has been accounted for by augmenting the 
noise),\footnote{For example, for uncorrelated (i.e. white)
detector noise and a Gaussian beam profile 
$$
n_\ell = n_0 \exp \left[ \ell(\ell+1) {\theta _{fwhm}}^2 / (8 \ln 2) \right] .
$$
}
and the absence of foregrounds, 
the following optimal estimator may be constructed
for $f_{NL}$ (assumed to be small)
\be
\hat f_{NL}^{ideal}=
\frac{
\langle b^{th}_{f_{NL}=1}, b^{obs} \rangle 
}
{
\langle b^{th}_{f_{NL}=1}, b^{th}_{f_{NL}=1} \rangle 
}.
\label{IdealEst}
\ee
Here the natural bispectral inner product (relative to a
particular experiment characterized by $n_\ell$) is defined as 
\be
\langle b^A, b^B \rangle =
 \sum _{\ell _1\le \ell _2 \le  \ell _3}
\frac
{
b^A_{\ell _1 \ell _2 \ell _3}
b^B_{\ell _1 \ell _2 \ell _3}
}
{{\rm Var}[b_{\ell _1 \ell _2 \ell _3}^{obs}]} ,
\label{InnerProd}
\ee
where the variance ${{\rm Var}[b_{\ell _1 \ell _2 \ell _3}^{obs}]}$ 
is calculated according to eqn.~(\ref{VarExp}) 
except that $c_\ell $ has been replaced by 
$(c_\ell +n_\ell ).$
[In the remainder of this paper ${\rm Var[b]}$ will always include
the noise $n_\ell$.]
In eqn.~(\ref{IdealEst}) 
$b^{obs}$ represents the observed bispectrum, which here is  
rescaled to correct for 
the attenuation from the finite width beam profile.
Note that we are here using the Gaussian assumption to calculate 
the variance, which is only correct in the limit of weak non-Gaussianity.
In fact, as we will see in section~\ref{validation} and as has been pointed 
out in \cite{creminelli, liguori}, 
if the value of $f_{NL}$ corresponds to a detection at several
sigma for a given experiment, non-Gaussian corrections to the variance
become important.

The estimator above can be rewritten in the form
\be
\hat f_{NL}^{ideal} = \sum_{\ell _1\le \ell _2 \le  \ell _3} 
w_{\ell_1 \ell_2 \ell_3} 
\frac{b^{obs}_{\ell_1 \ell_2 \ell_3}}
{b^{th (f_{NL}=1)}_{\ell_1 \ell_2 \ell_3}}, 
\ee
which makes apparent the fact that the optimal estimator
is an inverse variance weighted linear combination of
the estimators that would be obtained for each independent
$(\ell _1, \ell _2, \ell _3)$ triplet. 
To leading order, estimators corresponding to
distinct triplets are uncorrelated. 
The inner product 
$\langle b^{th}, b^{th} \rangle$
defined in (\ref{InnerProd}) is equal to the 
total $\chi ^2$ or $(S/N)^2$
for detecting a bispectral template 
$b^{th}_{\ell _1 \ell _2 \ell _3}$
for an experiment whose noise and beam profile
are defined by $n_\ell$.

The estimator presented here assumes uniform sky coverage.
However many scanning strategies, and in particular the one
adopted for Planck, cover some parts of the sky more
densely than others. For the present estimator in the form
described here, such non-uniform noise will lead to an increase
in variance but without any bias. Because of the non-uniformity
of the essentially white instrument noise, products of map pairs,
particularly those of large wavenumber, will be correlated with
the scan pattern, which includes power at low wavenumber. This
means that correlations with a third map of low wavenumber
(i.e., the triplet combinations contributing predominantly
to the total available signal-to-noise squared for local
bispectral non-Gaussianity) will include a product of
the low-$\ell$ projection of the scan pattern with the low-$\ell$
Gaussian component of the primordial anisotropy, the expectation value 
of which is zero because the scan pattern is not
correlated with the primordial anisotropy. This effect can
be removed by calculating the expectation value of the
product arising from the scan anisotropy and subtracting
it. The effect of non-uniform sky coverage is not necessarily
all harmful, because deeper coverage of some parts of the
sky provides additional information which in principle could
be exploited to obtain less noisy correlations between the
low-$\ell$ primordial anisotropy, on the one hand, and the
local high-$\ell$ power, on the other hand. The above problem
and essentially the same strategy for mitigating it was
first noted and implemented in \cite{wmapCrem}. 

\subsection{Qualitative nature of signal}
\label{HandWaving}

To understand intuitively where the statistical information for
the non-Gaussian signal predicted from inflationary models
is situated, it is
instructive to express $(S/N)^2$ as a simple integral derived
under the following
simplifying assumptions. We employ the flat sky approximation,
ignore the discreteness of the multipole
numbers, and set all combinatorial factors to one. We further assume
that 
$c_\ell 
\sim \ell ^{-2}$ 
(in other words, a power spectrum devoid of acoustic oscillations,
Silk damping, and a finite width for the last scattering surface).
We assume 
$\bb _{\ell _1 \ell _2 \ell _3}
\sim 
\left(
 \ell _1^{-2} \ell _2^{-2} 
+\ell _2^{-2} \ell _3^{-2} 
+\ell _3^{-2} \ell _1^{-2} 
\right) ,$ which follows from exactly analogous assumptions.
Taking into account only cosmic variance, 
we obtain the triple integral
\ba
\left( \frac{S}{N} \right) ^2
&\sim &\Omega _{sky}
\int d^2\ell _1
\int d^2\ell _2
\int d^2\ell _3
\, \delta ^2( \ellb _1 +\ellb _2 +\ellb _3 ).
\cr
&& \times 
\frac{
\left(
 \ell _1^{-2} \ell _2^{-2} 
+\ell _2^{-2} \ell _3^{-2} 
+\ell _3^{-2} \ell _1^{-2} 
\right) ^2 
}{
\ell _1^{-2} \ell _2^{-2} \ell _3^{-2} 
}  
\label{SNIntegral}
\ea 
The integral obviously should be cut off both at small 
$\ell $ (because the quadrupole is the lowest accessible
multipole) and at large $\ell $ as well, because the finite
resolution of the experiment becomes relevant and serves
as a cut-off to render the integral finite.
By simple power counting, one would obtain
$(S/N)^2 \approx \Omega _{sky}\ell _{max}^2,$ but
considering elongated triangle configurations, 
with $\ell  _1\ll \ell  _2, \ell  _3,$ one obtains
a logarithmic divergence factor, refining our estimate
to become (see also \cite{babichzaldar})
\be
\left( \frac{S}{N} \right) ^2
\approx \Omega _{sky}\ell _{max}^2 
\ln \left( \frac{\ell _{max}}{\ell _{min}} \right) ^2.
\ee
We suppressed the dependence on $f_{NL}$ which would add a
factor ${f_{NL}}^2.$
The above estimate, despite the crudeness of the approximations
employed, highlights a number of qualitative features of the 
non-Gaussian signal predicted from inflation. The presence of 
the logarithm emphasizes the importance of the coupling between
the largest and smallest accessible scales of the survey. 
The $\Omega _{sky}\ell _{max}^2$ is proportional to the total
number of useable modes (or equivalently pixels), 
which would suggest that most of
the statistically significant information lies on scales
at the limit of the resolution of the survey. If there
were no logarithmic factor (in other words, no infrared
divergence in the above continuum integral), little information
would be lost by dividing the full-sky map into a large number
of smaller maps and forming a consolidated estimator by
averaging the $f_{NL}$'s estimated from
each of these submaps. However doing so would entail throwing away
the logarithmic enhancement factor, whose presence indicates
the importance of the correlations between the largest and smallest
scales of the survey. If it were not for acoustic oscillations,
which reverse the sign of the predicted bispectrum with respect
to the simplistic model adopted above, one could construct a 
crude caricature of the ideal estimator by dividing the survey
in two maps, so that  
\be
T(\hat \Omega )=
T_{\textrm{low-}\ell }(\hat \Omega )
+
T_{\textrm{high-}\ell } (\hat \Omega ) ,
\ee
and considering the correlation 
\be
\int 
d\hat \Omega ~
T_{\textrm{low-}\ell }(\hat \Omega )
T_{\textrm{high-}\ell } (\hat \Omega )
T_{\textrm{high-}\ell } (\hat \Omega ),
\ee
whose expectation value would vanish in the Gaussian theory, and
which measures to what extent the 
$\textrm{high-}\ell $ power 
is modulated by the 
$\textrm{low-}\ell $ anisotropy pattern. 
We shall see later on that this estimator is far closer to the 
actual estimator than one might believe at first sight. 

\subsection{General linear $f_{NL}$ estimators}

The inner product defined in (\ref{InnerProd}) is useful for 
calculating
how much information is lost by employing a non-ideal
template to estimate $f_{NL}$ rather than the exact 
theoretical 
template appearing in (\ref{IdealEst}). Suppose that we use the
template 
$b^{templ}$
to form the estimator (where the factor in the denominator
provides the normalization needed to render the 
estimator unbiased)
\be
\hat f_{NL}^{non-ideal}=
\frac{
\langle b^{templ}, b^{obs} \rangle
}
{
\langle b^{templ}, b^{th}_{f_{NL}=1} \rangle
} .
\label{NonIdealEst}
\ee
In principle any template with a non-vanishing overlap
with the theoretical prediction could be used.
The lack of optimality is characterized by the ratio
of variances
\be
\frac{
\Var ( \hat f_{NL}^{non-ideal})
}{
\Var ( \hat f_{NL}^{ideal})
}
=
\frac{
\langle b^{th}_{f_{NL}=1}, b^{th}_{f_{NL}=1} \rangle
}
{
\langle b^{templ}, b^{th}_{f_{NL}=1} \rangle
}
=
\frac{1}{\cos ^2\theta } ,
\label{DeviationIdeality}
\ee
indicating by what factor the variance is increased relative 
to the minimum variance estimator. Here
$\theta $ is the angle between the directions determined by
the two templates in template space.

There are several reasons why using a non-ideal estimator
could be attractive. Firstly, to compute the ideal
estimator exactly would require an inordinate amount of 
computer time, most of which would be of marginal value
as the $\Delta \ell =1$ spectral resolution is not needed.
Secondly, in the presence 
of contaminants the above weighting scheme is no longer ideal anyway and 
another weighting must be employed. Thirdly, let us assume that a 
detection is made by means of a single number. Then we would want
to divide data in many different ways to make sure that
the detected signal is not of another type with fortuitous
overlap with our favorite theory. For a detection 
with significant signal-to-noise showing that subdividing
$\ell _1 \ell _2 \ell _3$ space in different ways
always gives a consistent $f_{NL}$ would lend credence
to a claimed detection.
These points will be investigated
in more detail below.

The ideal estimator is time consuming to calculate.
If we cut off the sum in harmonic space at, say, $\ell \sim 10^3,$ 
we need to consider $\sim 10^{9}$ reduced bispectrum terms 
(actually the true number is about a factor 10 
smaller because of the triangle inequality) and the calculation of a 
typical observed bispectrum coefficient would involve 
$\sim 10^{6}$ terms. 
KSW\citep{koma} discovered a clever algorithm for calculating
the above estimator, but this method is restricted to producing just
the single number $f_{NL}$, and has fixed weights. To go beyond these
restrictions another way to speed up the computation is desirable.

Obviously the data analysis challenge 
just presented is exaggerated, because the limit on
the spectral resolution is without relation to
physics underlying the CMB, and the scales
over which interesting variations in the predicted bispectrum
may be expected to occur are characterized by 
the period of the acoustic oscillations
$\Delta \ell \approx 200$, 
and the damping scale $\ell _{damp}\approx 400$, which is 
due both to Silk damping and the finite width of the last scattering
surface. It is not necessary to compute
the bispectrum for every $\ell$-triplet.
One can instead bin in harmonic space to drastically reduce the 
computational effort.

\subsection{Binning}

We define a coarse-grained, or binned, estimator by partitioning
into a sequence of $M$ bins
\be
{\cal I}_{\cal A}=[\ell _{\cal A}, \ell _{{\cal A}+1}-1]
\ee
where ${\cal A},{\cal B}=(1,\ldots, M)$ and
$ \ell _1 < \ell _2 < \ldots < \ell _{M+1}$
(the last bin includes $\ell_{M+1}$).
We define the binned bispectrum as 
\be
\tilde{b}_{ {\cal A} {\cal B} {\cal C}} =
\sum _{\ell _A\in {\cal I}_{\cal A}} 
\sum _{\ell _B\in {\cal I}_{\cal B}} 
\sum _{\ell _C\in {\cal I}_{\cal C}} 
b_ {\ell _{\cal A} \ell _{\cal B} \ell _{\cal C}} 
\label{DefBinned}
\ee
and an analogous expression for the binned variance 
${\rm Var}[\tilde{b}_{\cal ABC}]$.

The increase of the variance due to binning can be quantified using
the $\cos^2\theta$ defined above in (\ref{DeviationIdeality}).
If binning is used, it corresponds to the following bispectral template:
\be
b^{binned}_{\ell_1 \ell_2 \ell_3} = 
\frac{{\rm Var}[b_{\ell_1 \ell_2 \ell_3}]}
{{\rm Var}[\tilde b_{\cal ABC}]}
~ \tilde{b}^{th}_{{\cal A} {\cal B} {\cal C}} ,
\ee
where $({\cal I}_{\cal A},{\cal I}_{\cal B},{\cal I}_{\cal C})$ 
in this expression is that particular 
bin-triplet that contains the mode-triplet $(\ell_1,\ell_2,\ell_3)$. 
Results for different binnings are presented later
in the paper. Note that 
\be
\langle b^{binned}, b^{th} \rangle
= \langle b^{binned}, b^{binned} \rangle.
\ee 

\subsection{Dealing with bispectral contaminants}

In the presence of contaminants the above weighting
scheme is no longer ideal and another weighting must be employed
(see also \citep{smithzaldar}).
This situation may be illustrated by the following simplified
model, where the observed bispectrum may be decomposed
into the sum
\be
b^{obs}= f_{NL}b^{th}+ b^{cont}+ b^{noise}.
\ee
Here all the $b$'s have three subscripts 
$\ell _1,$ $\ell _2,$ $\ell _3,$
which have been suppressed to render the notation more compact, and 
$b^{cont}$ is a contaminant bispectral component (for example due to 
incompletely subtracted foregrounds or instrumental
effects) and $b^{noise}$ is the noise from a particular realization
of the Gaussian theory. In order to modify our estimator to mask
the effect of the contaminant, we must characterize the statistical
properties of the contaminant bispectrum. It is reasonable to assume
that 
\be 
\langle b^{cont} \rangle =0 ,
\ee 
because any known bias can be removed by direct subtraction
from the observed bispectral signal. The simplest statistical description for 
$b^{cont}$ would be a Gaussian, which would be completely characterized 
by the correlation function
\be
C^{cont}_{ \ell_1 \ell_2 \ell_3; \ell_1 ^\prime \ell_2 ^\prime \ell_3 ^\prime }
=\langle 
b^{cont}_{ \ell _1 \ell _2 \ell _3}
b^{cont}_{\ell _1 ^\prime \ell _2 ^\prime \ell _3 ^\prime }
\rangle .
\ee
We compute the likelihood ignoring constant factors not depending on $f_{NL}$
and integrating out the unknown contaminant $b^{cont}$
\ba
{\cal L} & \sim & \int (\mathcal{D}b^{cont})
\exp \left[ -\frac{1}{2}{b^{cont}}^T {(C^{cont}})^{-1} b^{cont}\right]
\cr &&
\times
\exp \biggl[ -\frac{1}{2} (b^{obs} -b^{cont} -f_{NL}b^{th})^T 
\cr &&
\qquad 
\times
({C^{G}})^{-1} 
(b^{obs}-b^{cont}-f_{NL}b^{th})\biggr] ,
\ea
which may be evaluated to (up to an irrelevant factor)
\be
\exp \left[ -\frac{1}{2} (b^{obs} -f_{NL}b^{th})^T (C^{G}+C^{cont})^{-1} 
(b^{obs} -f_{NL}b^{th})\right] ,
\ee
which we would like to put into the form 
\be
{\cal L} \sim \exp \left[-\frac{1}{2} \frac{(f_{NL}-\hat{f}_{NL}^{ml})^2}
{\sigma_{f_{NL}}^2}\right]. 
\ee
We obtain the following optimal estimator for this modified situation
\be
\hat f_{NL}^{ml}=
\frac%
{{b^{th}}^T (C^G+C^{cont})^{-1} b^{obs}}
{{b^{th}}^T (C^G+C^{cont})^{-1} b^{th}},
\ee
and the ``information'' is given by
\be
\frac{1}{\sigma _{f_{NL}}^2}=
{{b^{th}}^T (C^G+C^{cont})^{-1} b^{th}}.
\ee

\section{Description of the theoretical bispectrum}

We now turn to examining more quantitatively the character
of the predicted reduced bispectrum 
$\bb (\ell _1, \ell _2, \ell _3 )$ that results
from the ``local'' bispectral anisotropy as a result of 
the ansatz (\ref{ansatz}).
Plots of the primordial bispectrum for various shapes 
can be found in \cite{JamesPaul,shapeNGcamb}, both 2D slices and 3D,
but here we focus in detail on the local shape.
All simulations in the rest of this paper assume cosmological parameters from 
the WMAP five-year best fit \cite{wmap5} using WMAP data only.
In Sect.~\ref{HandWaving} we examined the signal-to-noise ratio
making the most brutal approximation so that all relevant
quantities could be expressed in terms of simple analytic
expressions. In particular, under the assumptions made
there, we found that 
\be
\bb _{\ell _1 \ell _2 \ell _3}
\sim
\left(
 \ell _1^{-2} \ell _2^{-2}
+\ell _2^{-2} \ell _3^{-2}
+\ell _3^{-2} \ell _1^{-2}
\right) .
\ee
We would like to study the shape of the local bispectrum more
carefully in order to understand the deviations from this 
simple functional form, which moreover for plotting is
extremely useful because it converts the bispectrum to
a quantity that is comparatively rather constant. Therefore
it is useful to define and plot the dimensionless quantity 
\be
{\cal B}_{\ell _1 \ell _2 \ell _3}=
\frac{\alpha^2 \, \bb_{\ell _1 \ell _2 \ell _3}
}{
\frac{1}{\ell_1(\ell_1+1)\ell_2(\ell_2+1)} 
+ \frac{1}{\ell_2(\ell_2+1)\ell_3(\ell_3+1)}
+ \frac{1}{\ell_3(\ell_3+1)\ell_1(\ell_1+1)}
}
\label{BispecRatio1}
\ee
where for purposes of normalization $f_{NL}=1$ has been assumed.
The normalization constant $\alpha$ has been chosen such that 
$c_\ell^{-1} = \alpha \, \ell(\ell+1)$ in the Sachs-Wolfe approximation 
$\Delta T/T = -(1/3)\Phi$ where $\Phi$ is the
primordial Newtonian potential.

Another choice (leading to a quantity that is even more nearly constant) 
is 
\be 
\bar {\cal B}_{\ell _1 \ell _2 \ell _3}=
\frac{ \bb _{\ell _1 \ell _2 \ell _3}
}{
c_{\ell _1} c_{\ell _2}+ c_{\ell _2} c_{\ell _3}+ c_{\ell _3} c_{\ell _1}
}.
\label{BispecRatio2}
\ee
Here using the actual power spectrum in the denominator
factors out the overall decay (i.e., due to Silk damping
and the finite width of the visibility surface). However,
this quantity has the disadvantage that the oscillations
in $\bar {\cal B}$ become intertwined with the oscillations
of the two-point power spectrum.

\begin{figure}
\includegraphics[width=9.0cm]
{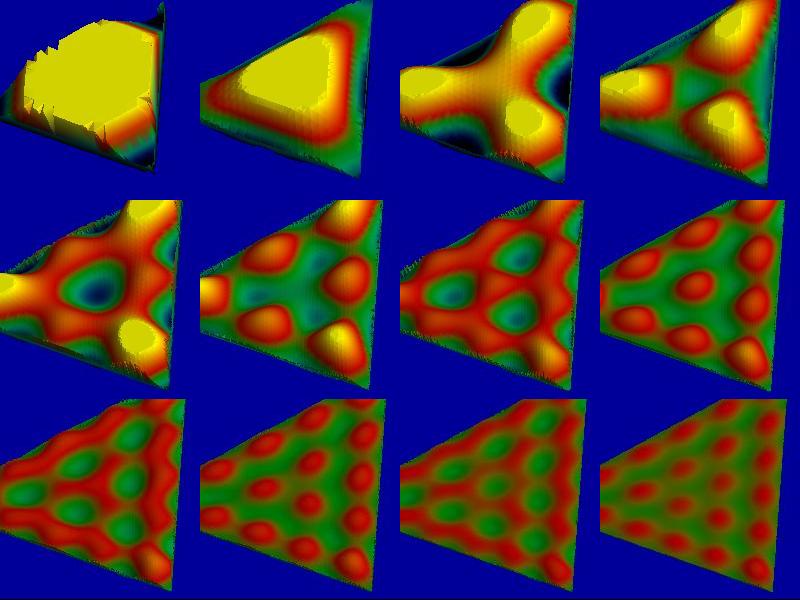}
\caption{We plot the reduced CMB TTT bispectrum rendered
dimensionless in a scale-free way according to the function
${\cal B}_{\ell _1 \ell_2 \ell _3}$
on 12 sections of constant 
$(\ell _1+\ell_2+\ell _3)/3$ corresponding to 
200, 300, 400, 500,
600, 700, 800, 900,
1000, 1100, 1200, 1300. The color scale (explained in the
text) ranges from $-14.0$ to $+14.0$ and values outside of this
range are clipped. 
}
\label{plate_one}
\end{figure}

\begin{figure}
\includegraphics[width=9.0cm]
{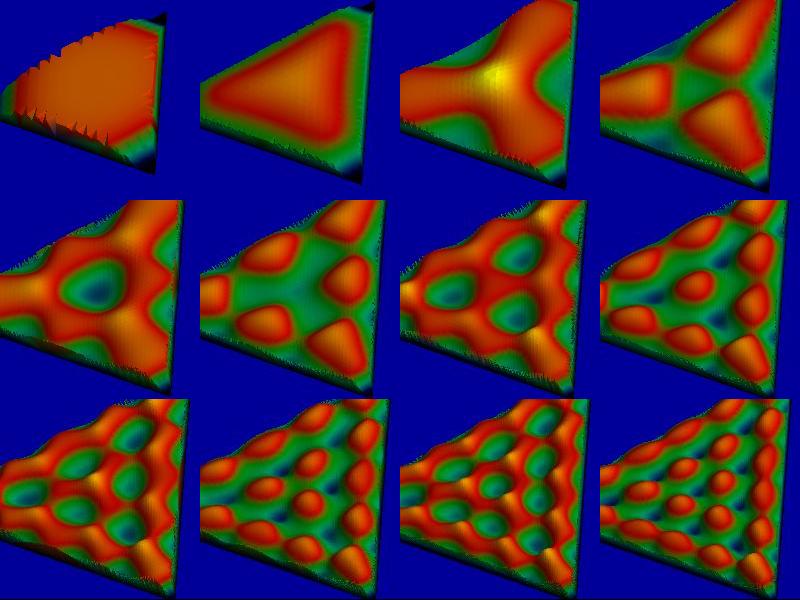}
\caption{Same as in Fig.~\ref{plate_one} except that
we instead plot $\bar {\cal B}_{\ell _1 \ell_2 \ell _3},$
normalized with the actual CMB temperature 
power spectrum rather than the scale
free one.
Here the color scale (same as above) ranges
from $-3.0$ to $+3.0.$
}
\label{plate_two}
\end{figure}

The panels of Fig.~\ref{plate_one} plot 
${\cal B}_{\ell _1 \ell _2 \ell _3}$,
showing a sequence of sections of constant 
$(\ell _1+\ell _2+\ell _3).$
Fig.~\ref{plate_two} shows the analogous plots for $\bar {\cal B}.$
The $xy$-coordinates in the plot correspond to the orthogonal
transverse coordinates
$\ell _{\perp a}=
(\ell _1-\ell _2)/\sqrt{2}$ and
$\ell _{\perp b}=(\ell _1+\ell _2 -2\ell _3)/\sqrt{6}.$ 
The function value is indicated by height
and color with the  color scale
ranging from $-14.0$ to $+14.0$ for ${\cal B}$ and
from $-3.0$ to $+3.0$ for 
$\bar {\cal B}.$
Black $\to $ blue $\to $ green indicates negative values
and
red $\to $ orange $\to $ yellow indicates positive values,
with the zero point situated between
green and red. 

The triangle inequality ($|\ell_1-\ell_2| \leq \ell_3 \leq \ell_1+\ell_2$ 
and cyclic permutations) reduces to
the following three conditions
\ba 
\qquad\qquad L_1 = \ell _2+\ell _3-\ell_1 &\ge &0,\cr 
\qquad\qquad L_2 = \ell _3+\ell _1-\ell_2 &\ge &0,\cr
\qquad\qquad L_3 = \ell _1+\ell _2-\ell_3 &\ge &0.
\label{TriIneq}
\ea 
Because the sum $(\ell_1+\ell_2+\ell_3)$ has to be even (parity condition),
the $L_i$ only take even values. The inverse relations are
\be
\ell_1 = \frac{L_2+L_3}{2}, \quad
\ell_2 = \frac{L_1+L_3}{2}, \quad
\ell_3 = \frac{L_1+L_2}{2}.
\ee
If all non-negative values of 
$\ell_1$, $\ell_2$, $\ell_3$
were allowed, the region 
allowed by the triangle inequality 
would comprise
the infinite triangular pyramid in the 
principal octant with $\ell _1,
\ell _2, \ell _3>0$
whose edges are the rays
$\ell_1=\ell_2,$ $\ell_3=0$
and the two other cyclic 
permutations thereof, and the faces would correspond
to $L_i=\textrm{(constant)}.$ The extremely elongated
triangles with one $\ell $ much smaller than the other two
are situated in the immediate neighborhood
of these edges. It is these configurations
that give rise to the logarithmic divergence previously
mentioned and contribute the bulk of the statistical
weight, in a way which we will characterize more quantitatively
later on.
Because the lowest 
observable multipole starts at
$\ell =2$ the edges are slightly truncated
by the intersecting planes $(\ell _1\ge 2,\ldots ).$
Fig.~\ref{plate_one} shows 16 sections of constant $(\ell_1+\ell_2+\ell_3)$.
The most negative values of ${\cal B}$ occur near 
the edges. In 
the central region we observe a sequence of acoustic
oscillations. 
First, we observe a single
maximum situation at the center of
the triangle (corresponding to an exactly equilateral
configuration). The center oscillates back down
but three new maxima appear situated between
the centre and the vertices.
The process repeats itself, so that the number of
local maxima progresses according to the sequence
$1,$ $3=1+2,$ $6=1+2+3,$ $10=1+2+3+4, ...$ 
[An animation of a sequence more finely sampled
is available at http://www.apc.univ-paris7.fr/$\sim$bucher/bispectrum.]

In Figs.~\ref{plate_three} and \ref{plate_four} we show isosurfaces of 
the function ${\cal B}_{\ell _1 \ell _2 \ell _3}$
at values -10, -4, -2, +2, +4, +10. The domain is an infinite triangular 
pyramid in the first octant and in the first panel all the isosurfaces 
for this most negative value terminate on the boundary. The long curved 
lens-like structures are situated along the edges and
track the acoustic oscillations of the two-point power spectrum, 
as discussed in more detail below. 
As the threshold isovalue is increased, new bubbles appear and coalesce.
The null isosurface is not shown because it has a regular Swiss cheese 
appearance, so that it is impossible to see inside. 

\begin{figure}
\begin{center}
\includegraphics[width=7.0cm]
{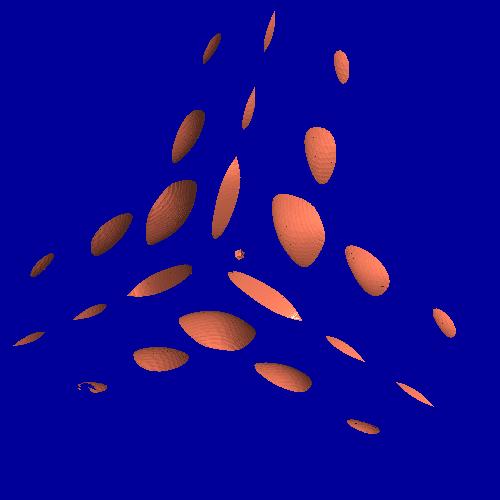}
\includegraphics[width=7.0cm]
{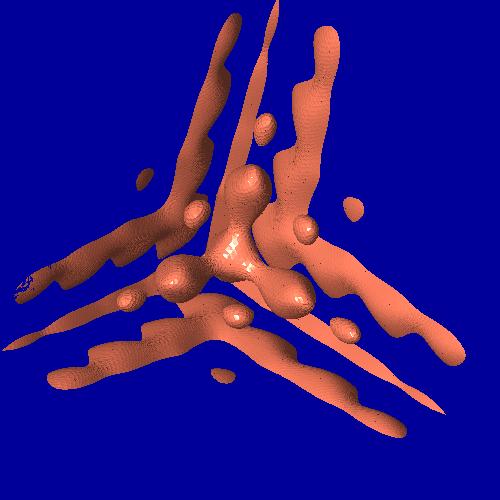}
\includegraphics[width=7.0cm]
{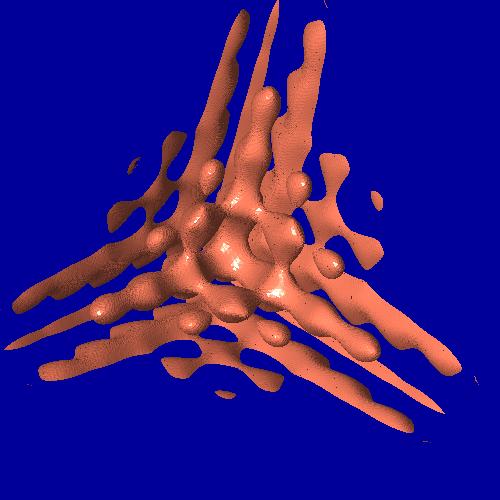}
\end{center}
\caption{
A sequence of isosurfaces for the function ${\cal B}_{\ell _1 \ell _2 \ell _3}$
is shown for ${\cal B}=$ -10, -4, -2.
[Additional and higher resolution isosurfaces may be found at the
website indicated in the text.]
}
\label{plate_three}
\end{figure}

\begin{figure}
\begin{center}
\includegraphics[width=7.0cm]
{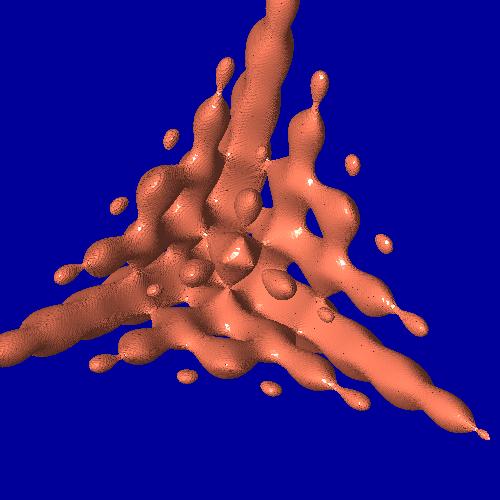}
\includegraphics[width=7.0cm]
{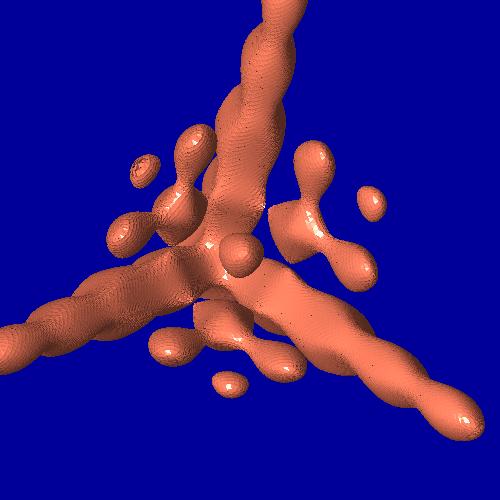}
\includegraphics[width=7.0cm]
{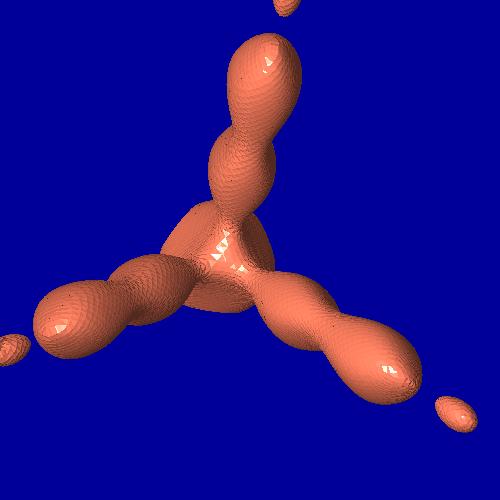}
\end{center}
\caption{Continuation of previous figure, with isosurfaces at ${\cal B}=$
2, 4, and 10.}
\label{plate_four}
\end{figure}

The TTT reduced bispectrum is given by \cite{acoustic}
\ba
\bb_{\ell _1 \ell _2 \ell _3} & = &
\vphantom{\int }
2 \, f_{NL} 
\left( \frac{2}{\pi} \right)^3 
\int _0^\infty {k_1}^2 dk_1 \int _0^\infty {k_2}^2 dk_2
\int _0^\infty {k_3}^2 dk_3~
\cr 
&& \times ~
\Delta _{\ell _1}(k_1) ~ \Delta _{\ell _2}(k_2) ~ \Delta _{\ell _3}(k_3)
\vphantom{\int }
\cr 
&& \times
\left[ P(k_1)P(k_2) + P(k_2)P(k_3) + P(k_3)P(k_1) 
\vphantom{\int }
\right] \cr 
&&\times
\int _0^\infty r^2dr ~
j_{\ell _1}(k_1r)~ j_{\ell _2}(k_2r)~ j_{\ell _3}(k_3r),
\label{RedBispec}
\ea
where the primordial power spectrum 
for the gravitational potential $\Phi$ 
is defined by
\be 
\langle \Phi(\mathbf{k}) \Phi^*(\mathbf{k'}) \rangle
= (2\pi)^3 \delta^3(\mathbf{k}-\mathbf{k'})P(k)
\ee
and the local ansatz (\ref{ansatz}) is assumed.
$\Delta _\ell (k)$ is the CMB transfer function
(computed with a Boltzmann solver, here using CAMB).
At the same time the expectation value for the CMB power spectrum is given by
\be
c_\ell = \frac{2}{\pi} \int_0^\infty k^2 dk \, P(k) \Delta_\ell^2(k).
\label{PowerSpec}
\ee

If we assume a simple power spectrum $P(k)$
(e.g., exact scale invariance
or a homogeneous power law), 
its contribution to the reduced bispectrum
is featureless and does not single out any special scales. 
However the CMB transfer function 
$\Delta _{\ell }(k)$
contains interesting physics characterized by 
a variety of scales (e.g., the scale
of matter-radiation equality, horizon size
at last scattering, the shape of the visibility
function, etc.) and it is these features, which
are presumably well understood, that should be 
helpful in distinguishing a primordial bispectrum
from other contaminants. 

We re-examine
some of the arguments previously put forth
by exploiting the approximate integral
for the statistical information in
eqn.~(\ref{SNIntegral})
in light of the exact local bispectrum
and exact inner product for a particular
experiment (here taken to be the PLANCK
instrument using an uncontaminated CMB map
obtained by combining in quadrature the
100, 143, and 217~GHz channels with the specifications
published in the PLANCK bluebook \citep{bluebook}).

\begin{figure}
\includegraphics[width=9.0cm]
{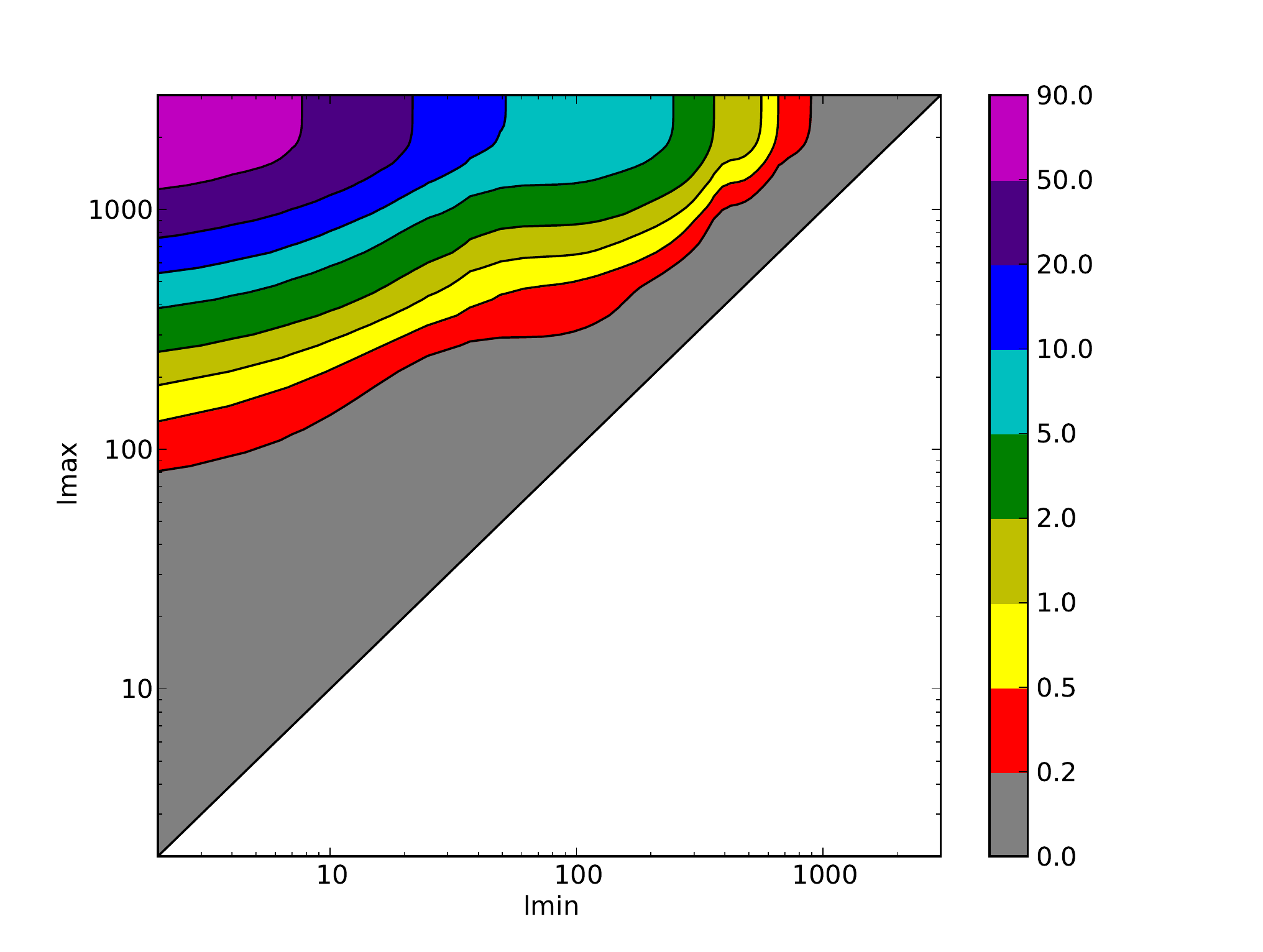}
\includegraphics[width=9.0cm]
{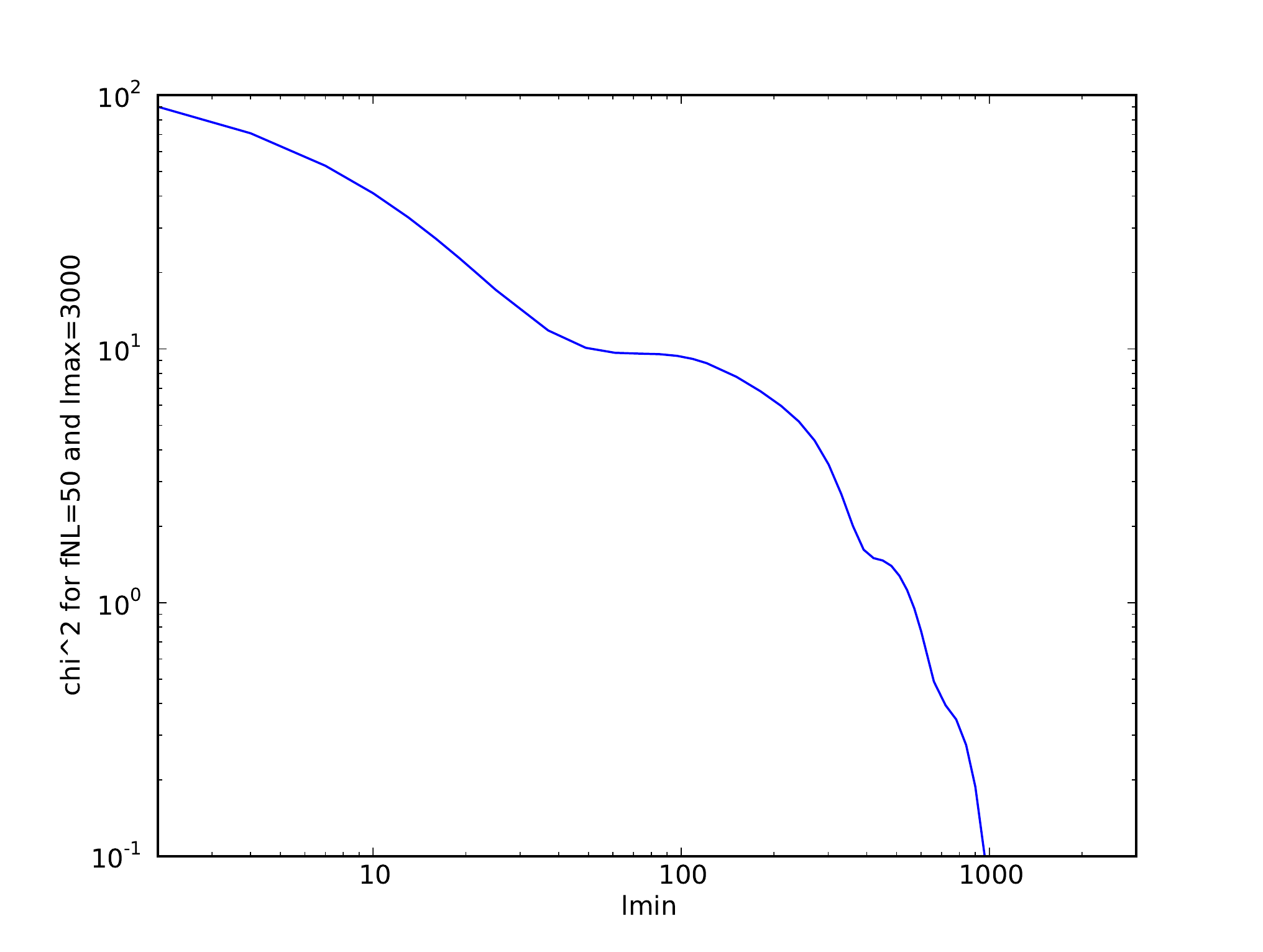}
\caption{We indicate the total $\chi ^2$ for a 
detection of $f_{NL}=50$ for the Planck experiment
taking into account the three frequency channels 100, 143, and 217~GHz.
In the upper plot $\chi^2$ is shown as a function of $\ell_{min}$ and
$\ell_{max}$ when only multipoles in the range
$\ell _{min} \le \ell \le \ell _{max}$ are taken into account.
In the lower
panel, a one-dimensional cut of the dependence on $\ell _{min}$ (while
$\ell _{max}$ is held at its maximum value of 3000) is shown.
}
\label{StatInfo}
\end{figure}

In order to visualize how the statistical weight of the signal is 
distributed in $\ell _1\ell_2\ell _3$ space, 
it is useful to make various cuts
and integrate $\chi ^2_{f_{NL}=50}$ taking into account these cuts.
(We take $f_{NL}=50$ as that is roughly the order of magnitude of
the central value of the WMAP5 results.)
The simplest cuts one can make is to restrict the multipole values to 
lie in a range $\ell_{min} \leq \ell \leq \ell_{max}$. The resulting
$\chi^2$ as a function of $\ell_{min}$ and $\ell_{max}$ is shown
in Fig.~\ref{StatInfo}.
We observe that almost all $(\approx 2/3 )$ the statistical weight
comes from the coupling of the lowest $\ell $ $(\ell \ltorder 10)$
with the largest $\ell $ $(\ell \gtorder 500)$.

\begin{figure}
\includegraphics[width=9.0cm]
{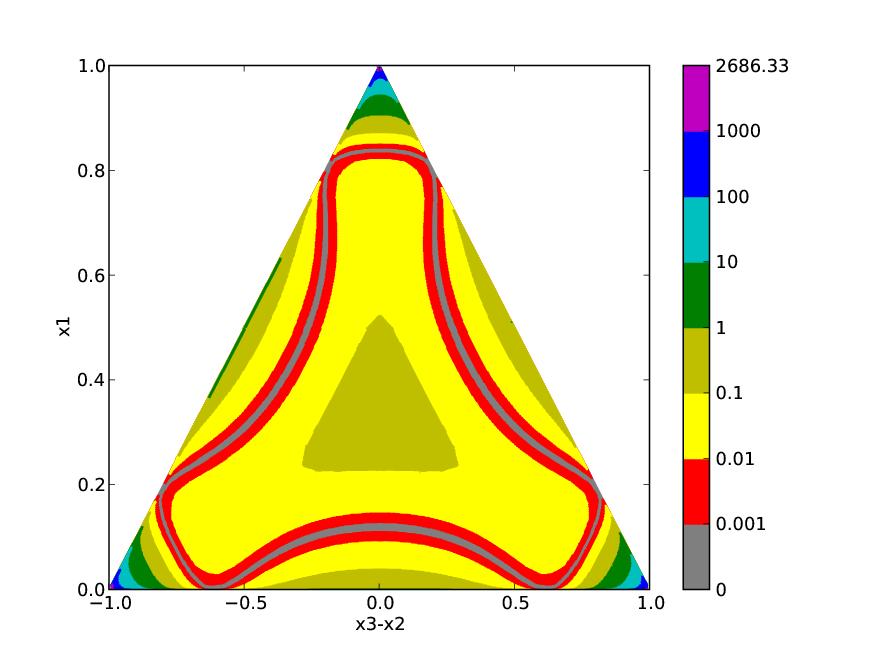}
\includegraphics[width=9.0cm]
{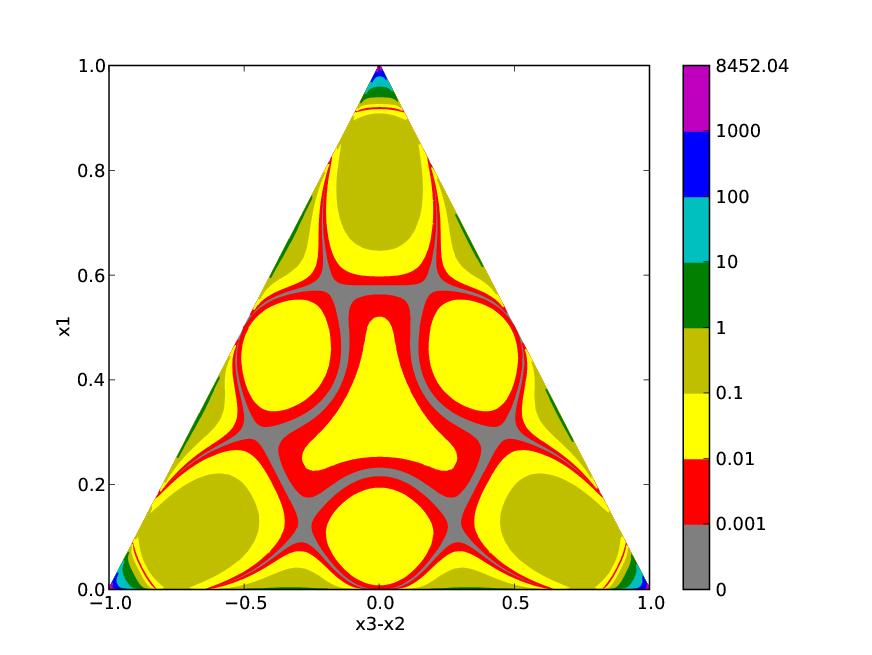}
\includegraphics[width=9.0cm]
{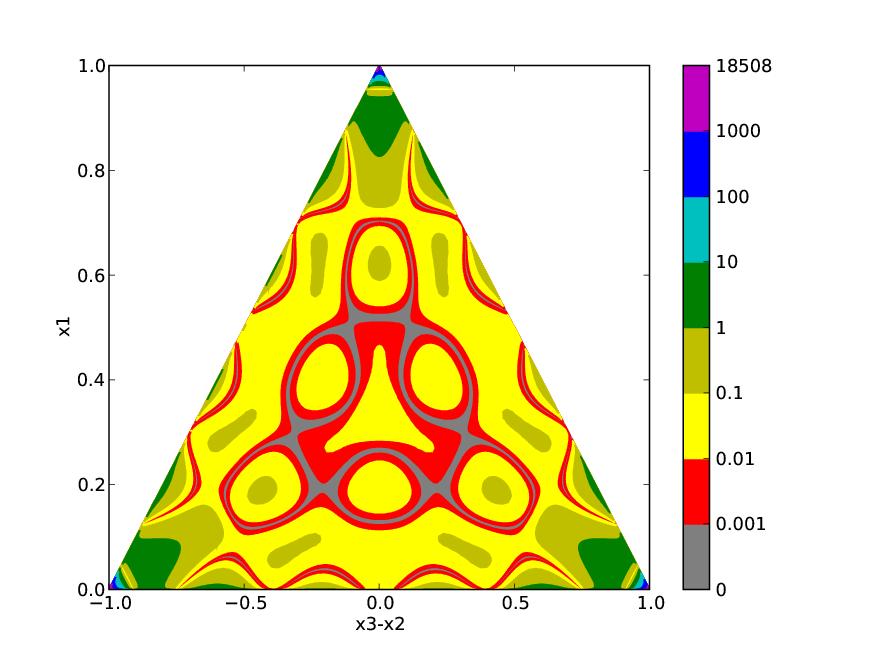}
\caption{To illustrate how the statistical
weight for a detection of $f_{NL}$ is distributed,
we plot the weights $(b^{th})^2/{\rm Var}[b^{obs}]$ 
(for the same three Planck channels as in the previous figure)
on three representative slices: 
$\ell _{sum}=1000$ (top),
$\ell _{sum}=2000$ (middle),
and $\ell _{sum}=3000$ (bottom). The normalization is such
that unity corresponds to the average value on the slice. 
In the central regions the grey contours correspond to sign changes
of the bispectrum. Note that the condition $\ell_1 \leq \ell_2 \leq \ell_3$
means that only 1/6 of the full triangle is relevant, the rest 
consisting of 5 identical reflected copies.
}
\label{WeightPlots}
\end{figure}

It is useful to adopt the following parameterization
under which there is one dimensionful quantity and a number of redundant
dimensionless quantities (of which only two are linearly independent). 
For our dimensionful quantity we choose 
$\ell_{sum}=\ell_1+\ell_2+\ell_3=L_1+L_2+L_3$. 
The dimensionless shape of the triangle in harmonic space is characterized by
\be
x_1=L_1/\ell_{sum}, \quad x_2=L_2/\ell_{sum},\quad x_3=L_3/\ell_{sum}
\ee
where $L_1, L_2,$ and $L_3$ are as defined in eqn.~(\ref{TriIneq}).
The quadruplet
$(\ell_{sum},x_1,x_2,x_3)$ provides a redundant set of coordinates 
(satisfying $x_1+x_2+x_3=1$) and the
triangle inequality is satisfied if and only if 
$0\le x_1\le 1,$ $0\le x_2\le 1,$ and $0\le x_3\le 1.$
The three edges of the allowed region are defined by 
$(L,1,0,0)$ and $(L,0,1,0)$ and $(L,0,0,1)$
where $L \geq 6.$

Fig.~\ref{WeightPlots} shows the weights $(b^{th})^2/{\rm Var}[b^{obs}]$ 
(the sum over $\ell_1, \ell_2, \ell_3$ of which corresponds to the $\chi^2$)
for three different slices with $\ell_{sum}$ fixed at 1000, 2000, 3000, 
respectively. 
Since the absolute value of the weights on
such a slice is not very meaningful, they have been divided by the average
value on the slice. In other words, values smaller than unity correspond
to regions with lower than average weight, and values larger
than unity to regions with higher than average weight.
The preponderance of the extreme configurations around the vertices 
is manifest in this figure, although their integrated contribution 
is not so easy to see.

\begin{figure}
\includegraphics[width=9.0cm]
{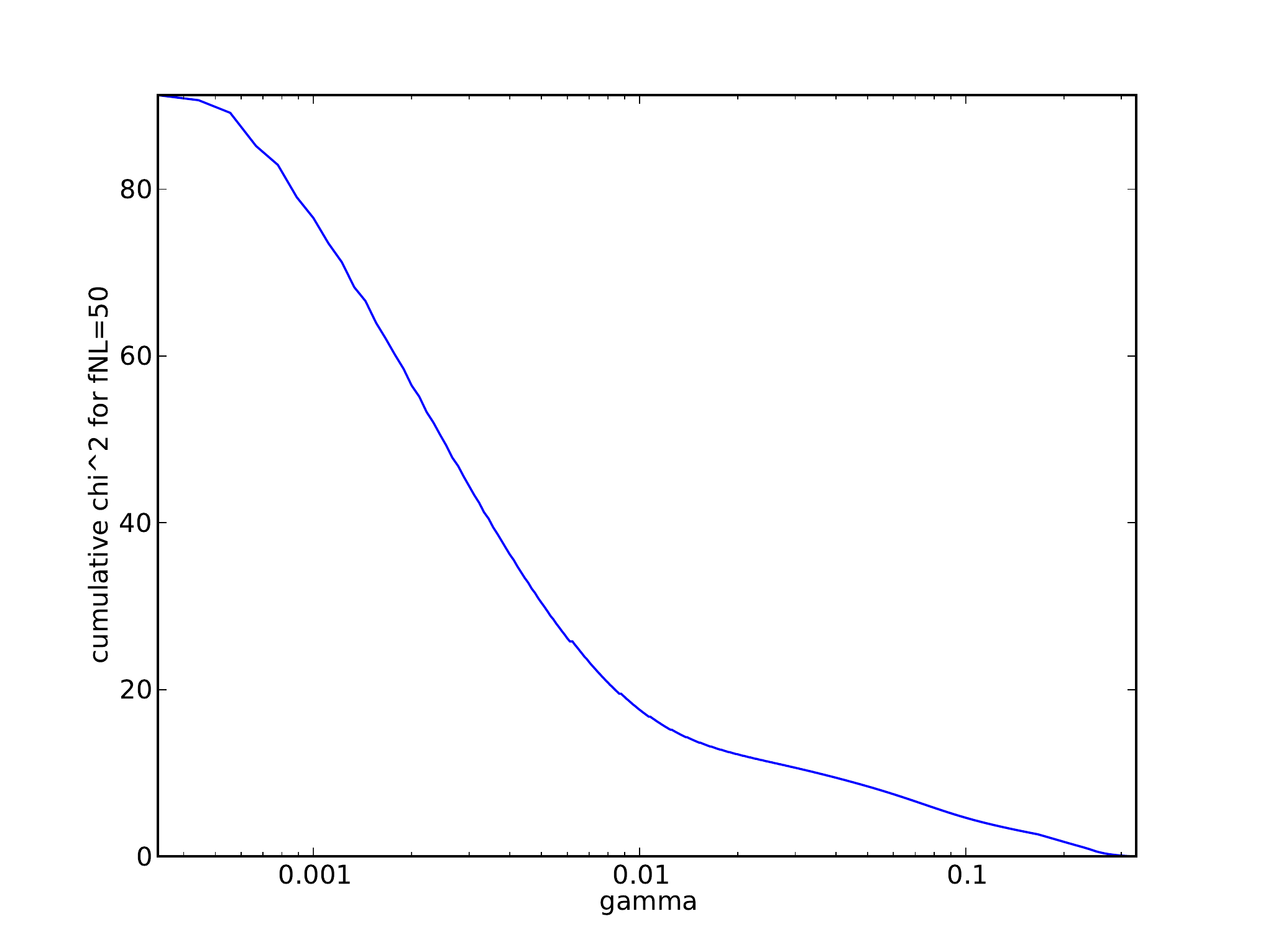}
\includegraphics[width=9.0cm]
{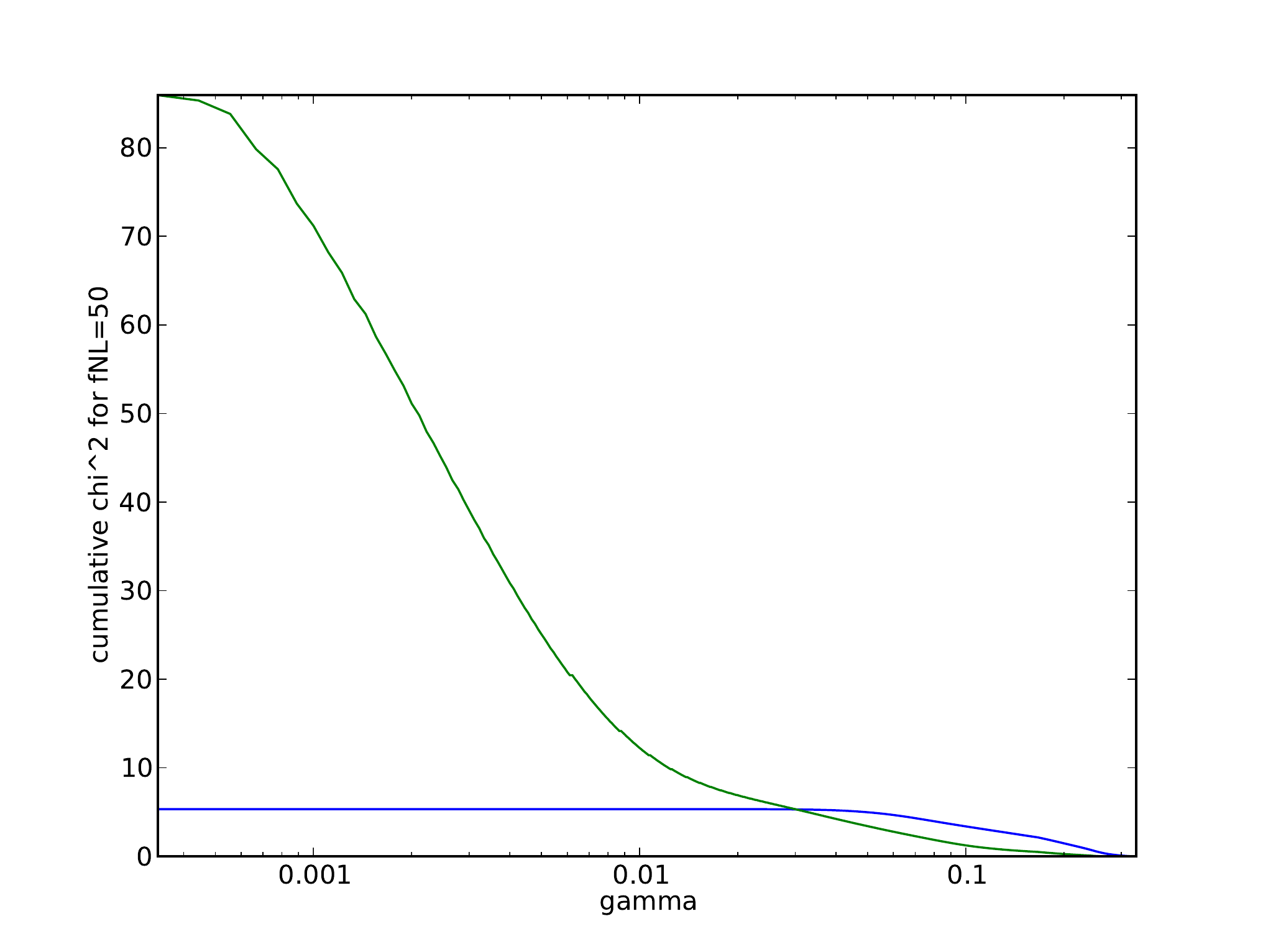}
\includegraphics[width=9.0cm]
{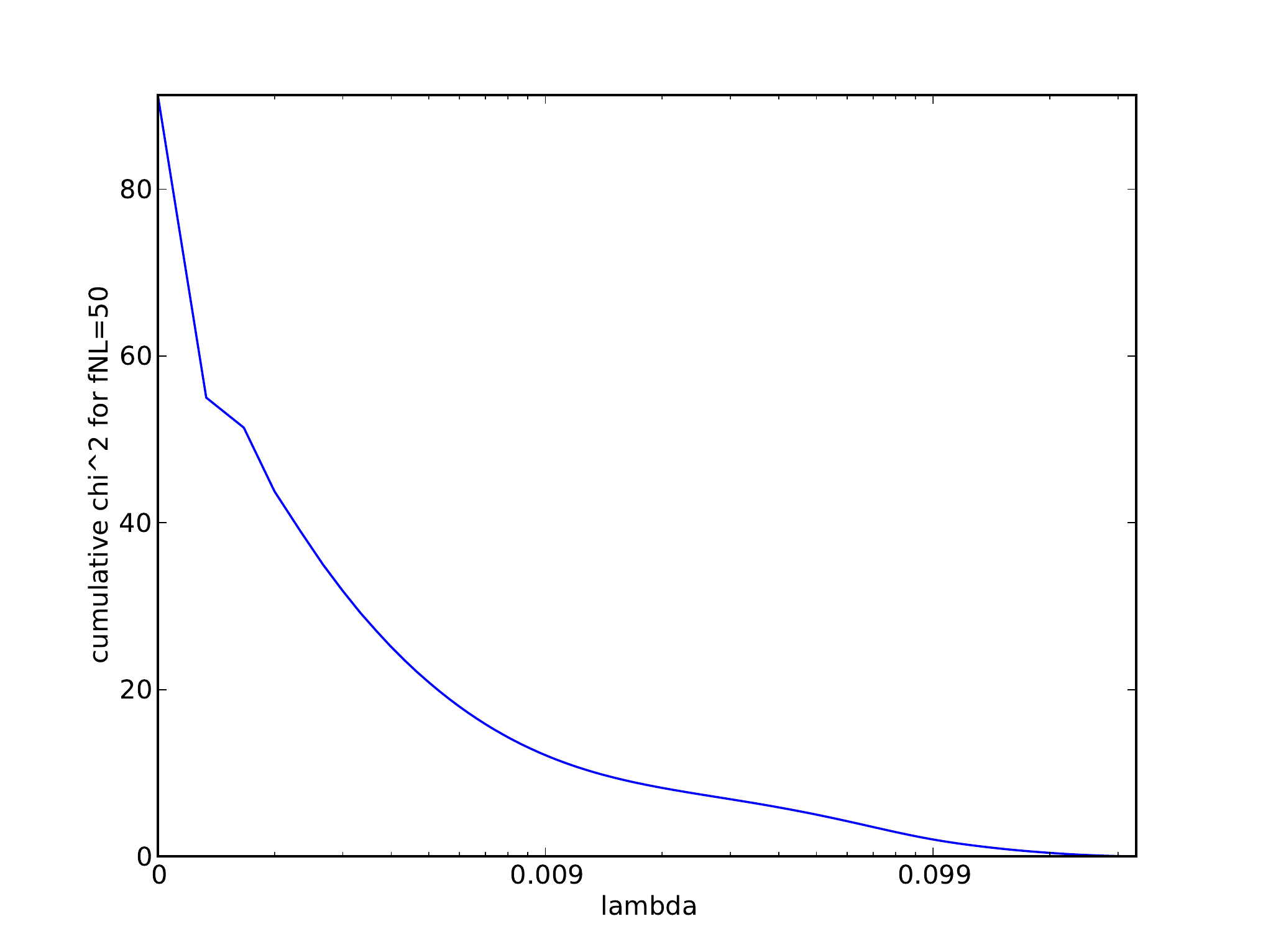}
\caption{The upper panel shows the effect on the total $\chi^2$
when one cuts away extreme triangles where one side is much 
shorter than the others. 
In the middle panel the contributions from positive (blue, the curve that
stays low) and negative (green) values
of the bispectrum are shown separately. The bottom panel shows
the effect of cutting away on the sides rather than the corners.
The minimum values of $\gamma$ and $\lambda$ (defined in the text)
correspond to taking all triangles into account, while for their maximum
value ($=1/3$) only equilateral triangles are taken into account.
The same three Planck channels as in the previous figures have been 
assumed, and $\ell_{max} = 3000$.
}
\label{CutCorner}
\end{figure}

We now define two families of cuts 
in order to quantify how much the triangles of the central region
contribute to the total integrated $f_{NL}$ signal. Firstly we may define
a family of expanding concentric equilateral 
triangles parameterized by $\lambda $ according to the relation
$\lambda \le x_1, x_2, x_3\le  (1- 2\lambda )$
with $0 \le \lambda \le 1/3.$
This is one way to cut out nearly flattened or collinear triangles.
Secondly, we may cut away at the corners
by requiring that 
$\ell_1/\ell_{sum}, \ell_2/\ell_{sum}, \ell_3/\ell_{sum} \geq \gamma $ 
where $2/\ell_{sum} \le \gamma \le 1/3$ so that the triangular region 
becomes hexagonal.
The results of the integrated $\chi ^2$ for $f_{NL}=50$ for these two families
of cuts is indicated in Fig.~\ref{CutCorner}. In the second panel the 
integrated
contributions for positive and negative values of the bispectrum are examined
separately. We observe that the exploitable signal is heavily concentrated
in the corners and very little statistically useful signal is present where
the bispectrum is actually changing sign.

\section{Prospects for detecting acoustic oscillations from vertex
configurations}

In the previous section we investigated the imprint of
physics occurring around the time of recombination on the
bispectrum, which intuitively may be characterized by
departures from a constant function of the bispectral ratio
${\cal B}_{\ell _1 \ell _2 \ell _3}$ defined in (\ref{BispecRatio1}). 
We observed
a rich pattern of acoustic oscillations in the
$x_1x_2x_3$-plane as $\ell $ varies. However
unfortunately, the statistical weight of this central
region is negligible, at least for an experiment like
PLANCK given the limits on $f_{NL}$ set by WMAP.
For PLANCK there is simply not enough signal-to-noise
in this region, as we have seen, although there may
be some hope with a next-generation satellite experiment
such as some of the most ambitious proposals for
a B-polarization satellite.

\begin{figure}
\includegraphics[width=9.0cm]
{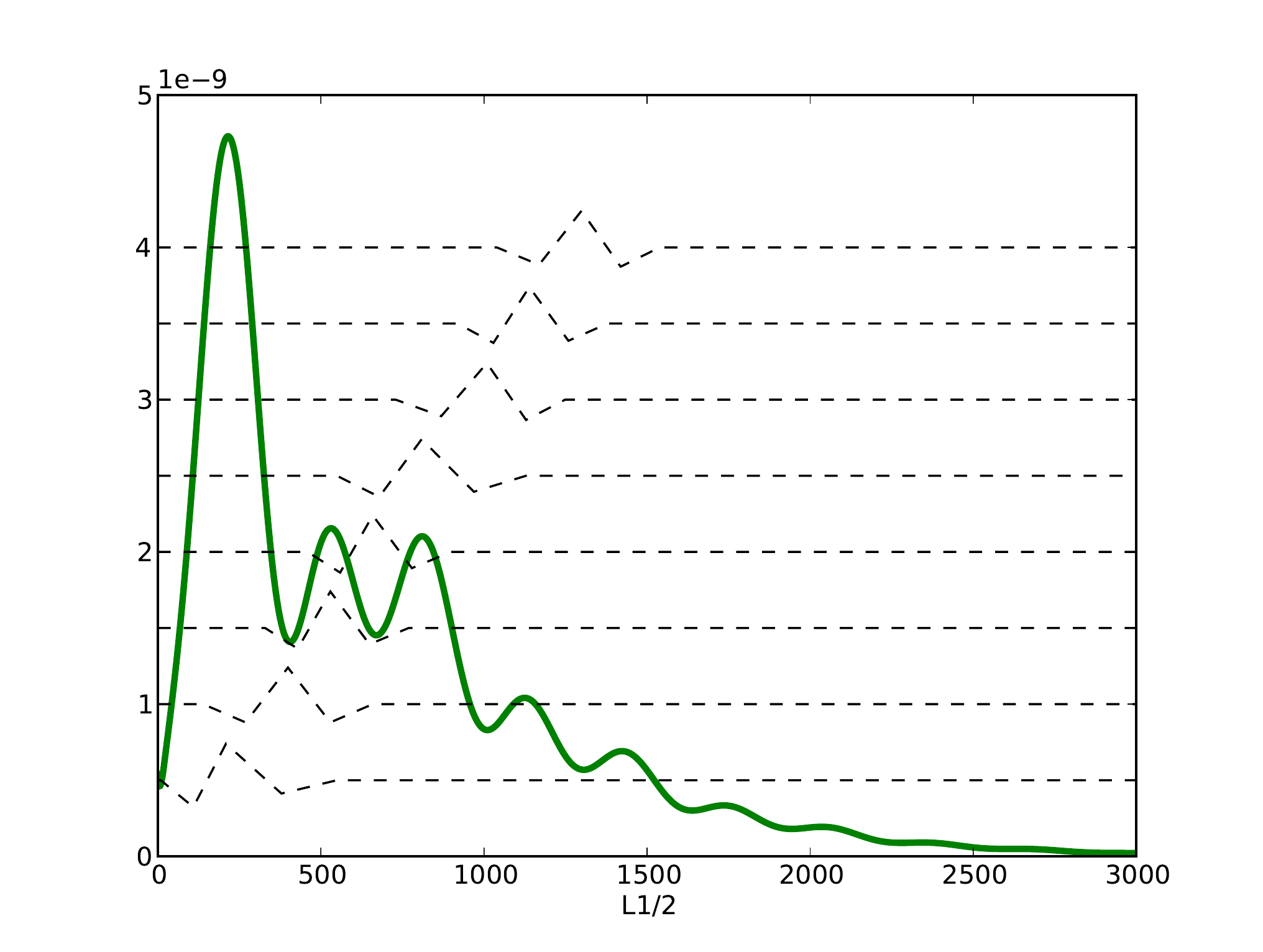}
\includegraphics[width=9.0cm]
{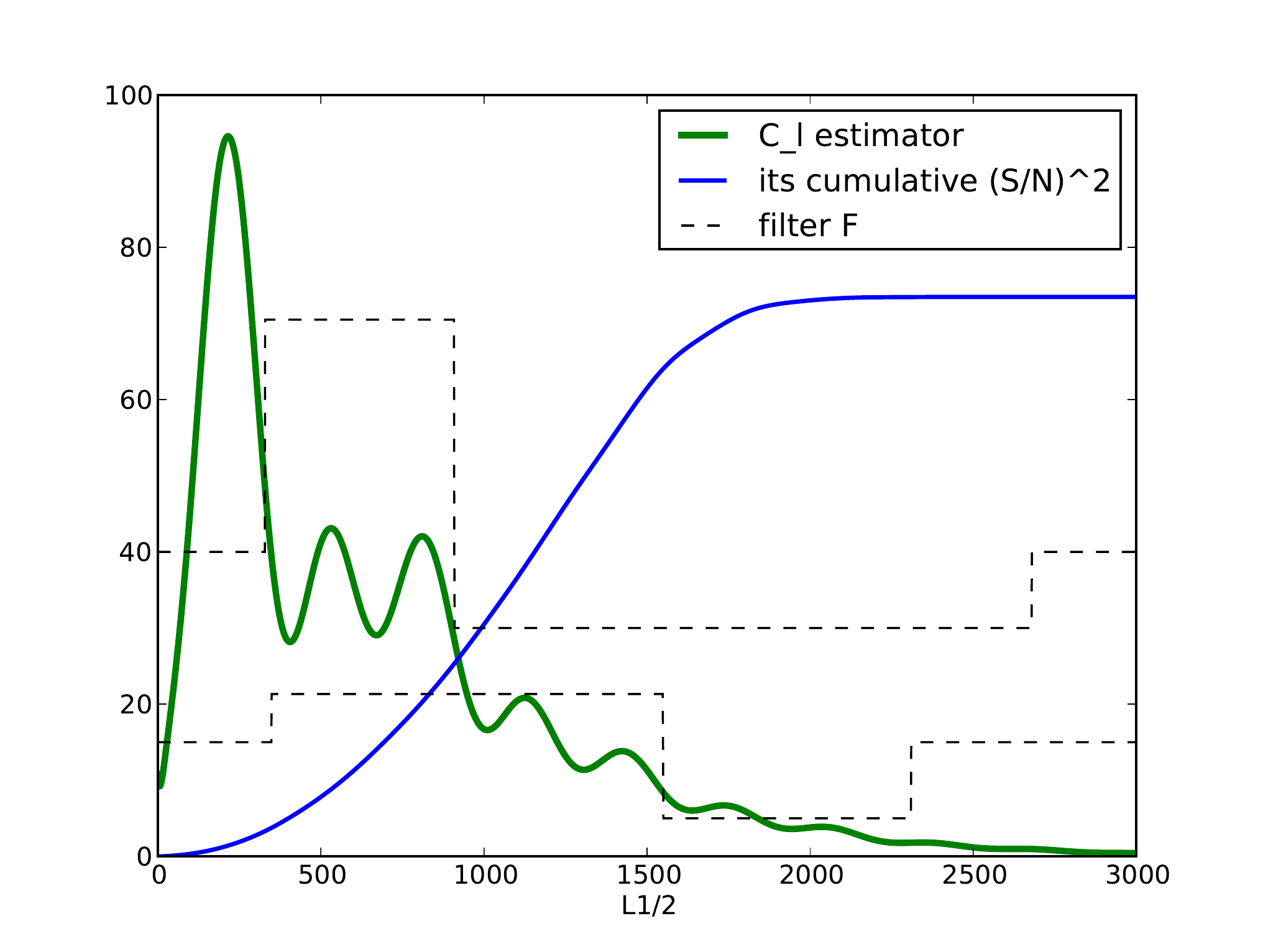}
\includegraphics[width=9.0cm]
{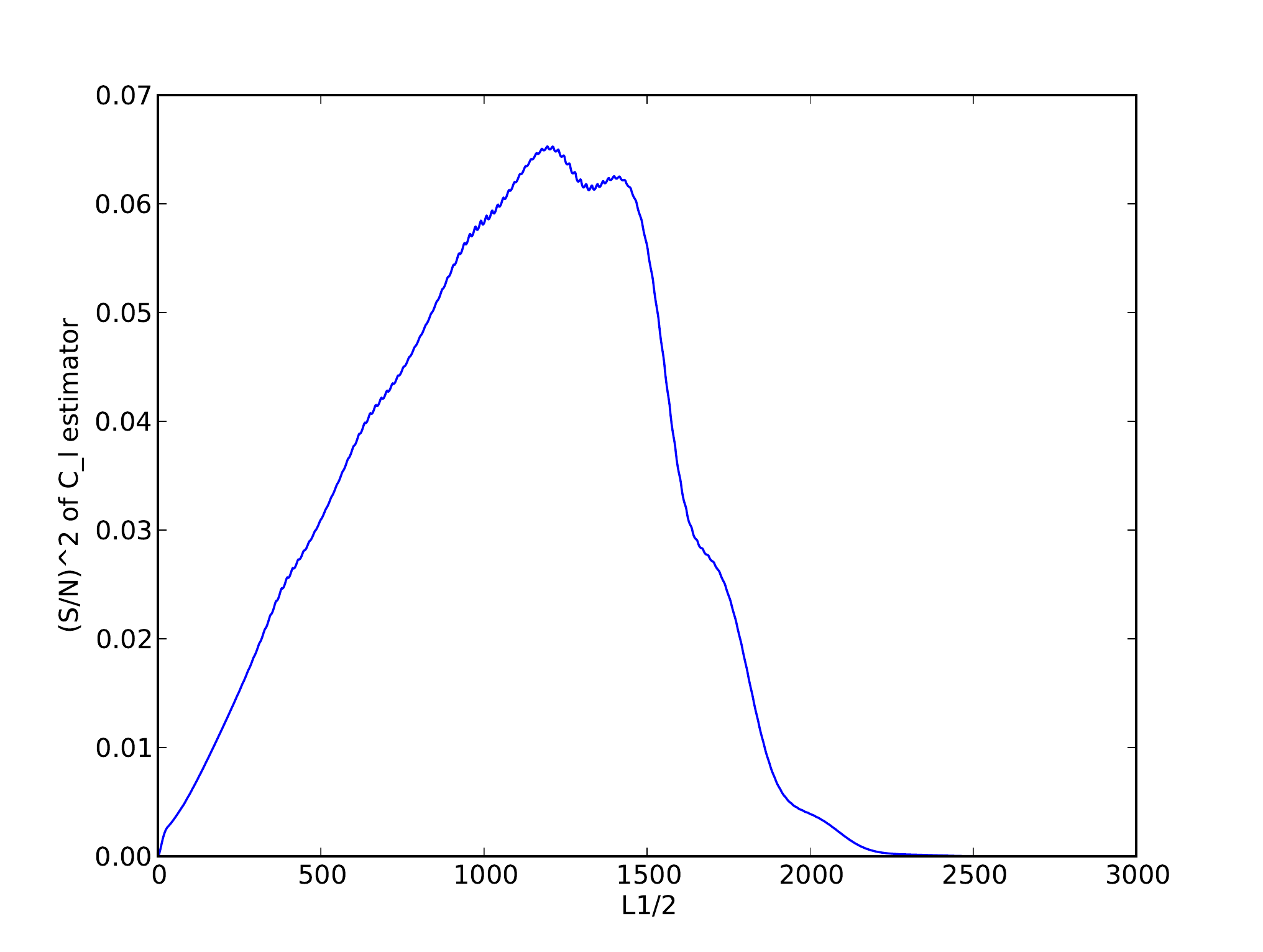}
\caption{
The bispectrum for triangles near the vertices of shape
space where $\ell _1\ll \ell _2 \approx \ell _3$ 
follows the shape of the two-point power spectrum as a function
of $L_1/2 \approx \ell_{large}$,
as shown in these plots of the estimator $\hat{\cal C}_{\ell_{large}}$
defined in (\ref{ClEstimator}) (green curves in the top and middle panels).
Its $(S/N)^2$ for an $f_{NL}=50$ detection is shown in 
the bottom panel, while the other solid curve (blue) in the middle panel 
shows the cumulative $(S/N)^2$ (and the values on the vertical axis of
this plot correspond to this curve). In the upper panel the dashed curves
show the shape of the discrete second-derivative filters used
to detect the peaks and the troughs. The two dashed curves in the 
middle panel show the first-derivative filters, which are much
broader, sensitive to the drop between low and high multipole
number. These filters are more leveraged and hence perform 
better. 
}
\label{LowLSection}
\end{figure}

The remaining window for detecting bispectral oscillations
resides in the $\ell$-dependence of $\bb _{\ell _1 \ell _2 \ell _3}$
in extreme vertex configurations. These configurations are defined
by $\ell _1\ll \ell _2, \ell _3$, which automatically means that 
$\ell_2 \approx \ell_3$ because of the triangle inequality. Hence
we define $\ell_{small} = \ell_1$ and 
$\ell_{large} = (\ell_2+\ell_3)/2 \approx \ell_2 \approx \ell_3$.
Using definition (\ref{BispecRatio2}), in this limit 
$\bb _{\ell _1 \ell _2 \ell _3}$
can be written as
\ba
\bb _{\ell _1 \ell _2 \ell _3} & = & \bar{\cal B}_{\ell _1 \ell _2 \ell _3}
\left( c_{\ell _1} c_{\ell _2} + c_{\ell _2} c_{\ell _3}
+ c_{\ell _3} c_{\ell _1} \right) \nonumber\\
& \approx & 2 \, \bar{\cal B}(\ell_{small}) \, c_{\ell_{small}} \,
c_{\ell_{large}}.
\label{SqueezedBispec}
\ea
Here we have used the fact (shown numerically) that for these squeezed 
triangle configurations $\bar{\cal B}$ is practically constant as a function
of $\ell_{large}$. We conclude that in the squeezed limit the bispectrum
is proportional to the power spectrum as a function of $\ell_{large}$.
This is shown explicitly in Fig.~\ref{LowLSection}.
[Alternatively, in a description that keeps the 3 different 
multipoles explicitly, one can use the $L_i$ defined in (\ref{TriIneq}). Then
the squeezed configuration corresponds with $L_2, L_3 \ll L_1$, 
the prefactor in (\ref{SqueezedBispec}) depends on $L_2$ and $L_3$ only, and 
we have a power spectrum behaviour as a function of $L_1/2$.]

To explain why $\bar{\cal B}$ does not depend on $\ell_{large}$ in the 
squeezed limit, we start from the integral expression (\ref{RedBispec}) 
for the bispectrum $\bb _{\ell _1 \ell _2 \ell _3}$.
In this limit the dominant contribution to the integral is proportional to
\ba
&&
\int _0^\infty \frac{dk_1}{k_1}\int _0^\infty {k_2}^2 dk_2
\int _0^\infty {k_3}^2 dk_3~
\left(
\frac{1}{{k_2}^3}
+\frac{1}{{k_3}^3}
\right)
\cr
&&\qquad \times ~
\Delta _{\ell _1}(k_1) ~ \Delta _{\ell _2}(k_2) ~ \Delta _{\ell _3}(k_3)
\vphantom{\int }
\cr
&&\times
\int _0^\infty r^2dr ~
j_{\ell _1}(k_1r)~ j_{\ell _2}(k_2r)~ j_{\ell _3}(k_3r),
\label{RedBispecBis}
\ea
where for clarity we insert an exactly scale invariant power
spectrum, with $P(k)\sim k^{-3},$
but all the arguments carry over 
to approximately scale invariant
spectra.
The dominant contribution to 
the radial integral in the last line occurs at $r\approx a$,
where $a$ is the radius of the surface of last scatter, and the
dominant values of the radial wave number are situated around
$k_i\approx \ell _i/a.$
The first spherical Bessel function varies slowly while the
the last two oscillate rapidly. This causes the integral to
cancel except near the resonance, where $k_2$ is very nearly equal 
to $k_3$ modulo a negligible offset of approximately $k_1.$ 
We therefore approximate 
eqn.~(\ref{RedBispecBis}) by
\ba
&&
\int _0^\infty \frac{dk_1}{k_1}\int _0^\infty {k_2}^2 dk_2
\int _0^\infty {k_3}^2 dk_3~
\left(
\frac{1}{{k_2}^3}
+\frac{1}{{k_3}^3}
\right)
\cr
&&\qquad \times ~
\Delta _{\ell _1}(k_1) ~ \Delta _{\ell _2}(k_2) ~ \Delta _{\ell _3}(k_3)
\vphantom{\int }
\cr
&&\times
\frac{\delta (k_2-k_3)}{k_2~k_3}~~
\int _0^\infty \frac{dr}{r^2} ~
j_{\ell _1}(k_1r)\cr
&\approx &
2 \int _0^\infty \frac{dk_1}{k_1}
\Delta _{\ell _1}(k_1) 
\int _0^\infty \frac{dr}{r^2} ~ j_{\ell _1}(k_1r)
\int _0^\infty \frac{dk_2}{k_2}
\Delta _{\ell _{large}}^2(k_2) 
\cr
&\approx &
\vphantom{\int }
f(\ell _1)~c_{\ell_{large}}.
\label{Factorization}
\ea
So we recover the fact that the bispectrum is proportional
to the power spectrum as a function of $\ell_{large}$, times a
prefactor that only depends on $\ell_{small}$.
We have confirmed by numerical integration of the 
exact expression that the factorization on the last line 
is a good approximation when $\ell_{small}$ is much less 
than the scales of the power spectrum acoustic 
oscillations and the damping. This 
factorization property greatly simplifies  
constructing estimators in the vertex region
by reducing the feature detection problem to 
essentially a one-dimensional problem.

We construct filters to detect the presence of several 
peaks and troughs in the bispectrum at positions corresponding
to the peaks and troughs of the two-point power spectrum,
which is of course now well known experimentally.
The mathematical problem of how best to construct these filters
is not completely well-defined,
so there is some arbitrariness in the choice of
filters designed to pick up the various features. Nevertheless
other filter choices can be expected to yield results differing
only slightly. We first describe our filter templates
as if they were acting on the two-point power spectrum
and then indicate how to carry over these templates
to the bispectrum problem, which is essentially a question
of how to deal with the two additional transverse dimensions.

We define the absence of features as the constancy of
${\cal C}_\ell=\ell (\ell +1)c_\ell$ and construct two types of filters.
The first takes a discretized second derivative of the power spectrum 
and is used to
detect the peaks and troughs. We require this filter
to give zero for a linear rise or fall in ${\cal C}_\ell$.
The second filter evaluates a discretized first derivative 
in order to detect a rise or fall in ${\cal C}_\ell$.

The first filter type evaluates 
an integral over some window function times the 
second derivative of 
${\cal C}_\ell $, where the support of the window function for 
a peak, for example,
would extend approximately from the trough on the left to the
trough on the right. This is the optimal width because making
the filter narrower would decrease the signal-to-noise by relying
on too few noisy data points and a broader filter
would bring in cancellations from neighboring regions of 
opposite sign, likewise diminishing the signal-to-noise.
One candidate for the second-derivative filter would
use the second derivative of a Gaussian profile, but 
this choice has
the drawback of an infinite support, which would have to
be cut off. Here we use instead a piecewise linear filter
profile.
For the first-derivative filter we use 
a step function constantly positive to the left of a certain value of
$\ell$ and constantly negative to the right. The central value, the width to
the left, and the width to the right are the three parameters in this filter.
The relative heights of the left-hand and right-hand sides are fixed by the
constraint that the integral of the filter should be zero.
The profiles used are shown in Fig.~\ref{LowLSection} in relation
to the predicted signal.

For the implementation in $ \ell _1 \ell _2 \ell _3 $-space,
we include only $\ell_{small}\le 30.$ Going any higher
would only marginally improve the signal and would tend to
smooth the power spectrum.
Then we group together triplets
with the same $\ell_{large}$. The function $f(\ell _{small})$
from the factorization in eqn.~(\ref{Factorization}) is found 
numerically and with each such group the best estimator for 
$\ell_{large}(\ell_{large}+1)\bb (\ell _{small},\ell_{large})/f(\ell _{small})$
(the expectation value of which is ${\cal C}_{\ell_{large}}$)
is constructed. The estimators for each triplet
are combined according to inverse variance weighting.
Explicitly, the resulting estimator $\hat{\cal C}_{\ell_{large}}$ is given by
(using the more precise notation $L_1/2$ instead of $\ell_{large}$)
\be
\hat{C}_{L_1/2} = N\left(\textstyle\frac{L_1}{2}\right)
\hspace{-0.4cm} \sum_{\begin{array}{c} 
\scriptstyle L_3 \leq L_2\\
\scriptstyle 2 \leq \ell_1 \leq 30
\end{array}} \hspace{-0.4cm}
\frac{2 \, \bar{\cal B}_{\ell_1 \ell_2 \ell_3}  
N_\triangle(\ell_1, \ell_2, \ell_3) \, c_{\ell_1}}
{\frac{L_1}{2} \left( \frac{L_1}{2} + 1 \right)
{\rm Var}[b^{obs}_{\ell_1 \ell_2 \ell_3}]}
\, b_{\ell_1 \ell_2 \ell_3} ,
\label{ClEstimator}
\ee
where $L_1 = \ell_2+\ell_3-\ell_1$ as before and 
the sum is over the two directions transverse to $L_1$, i.e. $L_2$
and $L_3$, under the condition that $\ell_1 \leq 30$. Here $N$ is the 
normalization of the inverse variance weights, given by
\be
\left[N\left(\textstyle\frac{L_1}{2}\right)\right]^{-1} 
= \hspace{-0.4cm} \sum_{\begin{array}{c} 
\scriptstyle L_3 \leq L_2\\
\scriptstyle 2 \leq \ell_1 \leq 30
\end{array}} \hspace{-0.4cm}
\frac{\left( 2 \, \bar{\cal B}_{\ell_1 \ell_2 \ell_3}  
N_\triangle(\ell_1, \ell_2, \ell_3) \, c_{\ell_1} \right)^2}
{\left(\frac{L_1}{2} \left( \frac{L_1}{2} + 1 \right) \right)^2
{\rm Var}[b^{obs}_{\ell_1 \ell_2 \ell_3}]} .
\ee
This estimator and its signal-to-noise ratio squared for the case of an
$f_{NL}=50$ detection are shown in Fig.~\ref{LowLSection}. It represents
one way of going beyond the single number $f_{NL}$ to describe the bispectrum
(see also \cite{munshia}).

To look for specific features, the estimator $\hat{\cal C}_{\ell_{large}}$
is summed with respect to the filters ${\cal F}(\ell _{large})$
described above. The resulting statistic $S$ is given by
\be
S = \sum_{L_1}
{\cal F}\left(\textstyle\frac{L_1}{2}\right)
\hat{C}_{L_1/2} .
\label{PeakEstimator}
\ee
The variance of $S$ is given by
\be
{\rm Var}[S] = \sum_{L_1} {\cal F}^2\left(\textstyle\frac{L_1}{2}\right)
N\left(\textstyle\frac{L_1}{2}\right).
\ee
The signal-to-noise ratios of this estimator in the case of $f_{NL}=50$
are given in Table~\ref{TableOne}.
Once more we have assumed the three Planck channels at 100, 143, and 
217~GHz, and we have used $\ell_{max}=3000$.
For an arbitrary $f_{NL}$ the table entries in the last column
should be multiplied by a factor of $(f_{NL}/50)^2$. 

\begin{table}
  \begin{tabular}{lllll}
  \hline
   Feature  & Center  &  $(\Delta \ell )_{left}$ & $(\Delta \ell )_{right}$ & $(S/N)^2$\\
 \hline
  1st peak  & 210 & 100 &  170 & 0.43  \\ 
  1st trough& 400 & 130 &  130 & 0.69  \\
  2nd peak  & 530 & 100 &  120 & 0.19  \\
  2nd trough& 660 & 100 &  120 & 0.27  \\
  3rd peak  & 810 & 130 &  160 & 0.70  \\
  3rd trough& 1010 & 140 & 120 & 1.08  \\
  4th peak  & 1140 & 110 & 120 & 0.34  \\
  4th trough& 1300 & 130 & 120 & 0.53  \\
\hline
  Drop A    & 910  & 580 & 1770 & 11.3 \\
  Drop B    & 1550 & 1200 & 760 & 21.9 \\
\hline
\end{tabular}
  \caption{Signal-to-noise squared for detecting 
  certain features in the edge bispectrum 
  $\ell _1\ll \ell _2, \ell _3$ for $f_{NL}=50.$ 
  Here we have cut with $\ell _1\le 30.$ The first
  eight filters are second derivative filters designed
  to detect peaks and troughs. Drops A and B are
  broader filters representing a discretized approximation
  of the first derivative centered between the third peak and third
  trough and between the fifth peak and fifth trough, respectively.
  These first derivative filters give a much larger $(S/N)^2$ 
  because of their extreme breadth. See Fig.~\ref{LowLSection}
  for the filter profiles.
}
\label{TableOne}
\end{table}

\section{Binned bispectral estimator implementation}

In this section we describe how to estimate $f_{NL}$ from a CMB map 
having been provided the details of the CMB experiment 
(instrument noise, beam profile, \ldots) and a
theoretical model, from which we require the primordial 
power spectrum $P(k)$ and transfer functions
$\Delta_l(k)$. As explained in a previous section, we need to evaluate
the binned approximation to 
the ideal estimator (\ref{IdealEst}) or to some variant 
(\ref{NonIdealEst}). We can write
the binned estimator as
\be
\hat{f}_{NL} = \sum_{{\cal A} \leq {\cal B} \leq {\cal C}}
\tilde{w}_{\cal ABC}(b^{th}_{f_{NL}=1}) \, \tilde{b}^{obs}_{\cal ABC}
\label{fNLestimator}
\ee
with the binned weights given by 
\be
\tilde{w}_{\cal ABC}(b^{th}_{f_{NL}=1}) = 
\frac{\tilde{b}^{th (f_{NL}=1)}_{\cal ABC}
/ {\rm Var}[\tilde{b}_{\cal ABC}]}
{{\displaystyle \sum_{{\cal D} \leq {\cal E} \leq {\cal F}}}
\left( \tilde{b}^{th (f_{NL}=1)}_{\cal DEF} \right)^2
/ {\rm Var}[\tilde{b}_{\cal DEF}]}.
\ee
Since we have changed the sum over $\ell_1 \leq \ell_2 \leq \ell_3$ into
a sum over $\cal A \leq B \leq C$, we have to divide the weights by
2 (or 6) in the case where 2 (or 3) bins are equal, to avoid overcounting.

Computing this estimator involves four distinct steps:
\begin{enumerate}
\item Computing the theoretical bispectrum $b^{th}_{\ell_1\ell_2\ell_3}$;
\label{step1}
\item Deriving the binned weights $\tilde{w}_{\cal ABC}$ from it;
\label{step2}
\item Computing the binned bispectrum $\tilde{b}^{obs}_{\cal ABC}$ from the map;
\label{step3}
\item Combining the above to yield $\hat{f}_{NL}$ (or another estimator).
\label{step4}
\end{enumerate}
The modular code we developed follows these 4 steps and the results from
each step are saved separately. Hence step~\ref{step3}, for example, 
is completely independent of steps~\ref{step1} and \ref{step2}, so that
when trying out another mask, say, only the code corresponding to 
steps~\ref{step3} and \ref{step4} has to be rerun.
While we focus in this section on $f_{NL}$, the modular setup of
the code makes it very easy to compute another bispectral quantity instead.
For example, by replacing in step~\ref{step2} the $f_{NL}$-weights defined 
above by the
binned version of the weights defined in (\ref{PeakEstimator}), we can
compute the statistic $S$ to detect acoustic bispectrum features.

\subsection{Implementation notes}

After getting the transfer functions from CAMB, to save computing time 
in step~\ref{step1}, eqn.\ (\ref{RedBispec}) for the reduced bispectrum 
is not evaluated for all $\ell$-triplets, but only for a representative
(large) number. The other necessary values are determined by 3D linear
interpolation. Our code also allows for the use of a 3D cubic 
interpolation scheme, but for the large number of exact ``grid points'' 
used we found no significant increase in accuracy compared to
the linear scheme, which is faster and already very accurate. 

To compute the weight for a bin in step~\ref{step2} one can either 
evaluate all terms inside the bin and sum them explicitly or,
much faster, for large bins use a simple 3D integration scheme
(here based on tetrahedronisation of the bin-cubes).
Both methods are implemented in our code. In the first case
all the values of $c_\ell$, $\bar{\cal B}_{\ell_1 \ell_2 \ell_3}$, and 
$N_\triangle(\ell_1,\ell_2,\ell_3)$ are pre-computed and saved to speed
up any consequent calculations for the same cosmological model
(at the cost of additional memory usage).
However, because of the slow variation of $\bar{\cal B}_{\ell_1 \ell_2 \ell_3}$,
the second (approximate) method is highly accurate and due to its
higher speed and lower memory usage generally to be preferred.

Computing the binned observed bispectrum in step~\ref{step3} involves 
multiplying the observed
$a_{\ell m}$ with the inverse beam and window function operators, applying
the masks (if any), and determining the maximally filtered maps 
$T_\ell(\Omega)$ according to (\ref{TempMap}). However, it is in this 
step that binning is a necessary ingredient to reduce computing time, 
so instead of 
computing the maximally filtered maps for each individual $\ell$, they are
only computed per bin, producing $T_{\cal A}(\Omega)$. Finally the binned
observed bispectrum $\tilde{b}^{obs}_{\cal ABC}$
is computed using the binned analogue of eqn.~(\ref{BispecFromMap}). 
Healpix \citep{healpix} is used to carry out the spherical harmonic transforms.

Using the files with the results from steps \ref{step2} and \ref{step3}
it is simple and quick to evaluate in step~\ref{step4} the sum 
(\ref{fNLestimator}) and find the result for $f_{NL}$. Its variance
is also computed. In the case of multiple maps (different realizations
of the same CMB sky) the result is computed in a number of different 
ways to check consistency: by first combining the different bins to
find a result per map, and then combining the different maps; or by first
combining the maps to find an averaged result per bin, and then combining
the bins. Moreover, using the set of maps we can also compute
an observed variance and use that in the weights instead of the theoretical
variance. In particular this allows us to see the effects of
deviations from Gaussianity in the variance, which become important
when $f_{NL}$ is several standard deviations away from zero for the
experiment under consideration.

Steps~\ref{step2} and \ref{step4} are fast (about 10 minutes and 10
seconds, respectively, on a laptop with an Intel Core 2 Duo P8400 processor 
(using only one core)), while step~\ref{step1} is slower (about 1.5 hours), 
all for $\ell_{max}=3000$. If necessary, step~\ref{step1} can be 
made faster by sampling the (at present probably unnecessarily densely 
sampled) theoretical bispectrum more sparsely.
Step~\ref{step3} is the slowest, taking about 45 minutes for one map with 
$\ell_{max}=2000$ and Healpix parameter $n_\mathrm{side}=1024$. However, we 
have parallelized this step, so that multiple maps can be run in parallel,
and further code optimizations are in progress. We run this code on a blade 
server with 16 quad-core Intel Xeon 5345 processors on 8 blades 
(64 cores in total with 4~GB of memory per core; one core of these performs
roughly the same as a core of the P8400 mentioned above), so that a set of 
100 maps takes about 1.5 hours.

\subsection{Binning scheme}

To run our code we have to specify a binning scheme. We choose a divisional 
scheme, although other schemes are implemented as well.
This means that we first split the interval $[\ell_{min}, \ell_{max}]$ into
two at the intermediate value $\ell_{int}=\ell_{min}+a_0(\ell_{max}-\ell_{min})$,
where the parameter $a_0$ lies between 0 and 1. Next each of these intervals
is split into two again using parameters $b_0$ and $b_1$, each rescaled to 
lie between 0 and 1. This process can be repeated as many times as necessary.

We can quantify how much the variance increases due to binning, compared
with an ideal estimator without binning, using the parameter
$\cos^2 \theta$ defined in (\ref{DeviationIdeality}). The results are 
summarized in Table~\ref{TableThree}. To find the maximum value
of $\cos^2 \theta$ we have varied the 7 parameters $a_0$, $b_0$, $b_1$, $c_0$,
$c_1$, $c_2$, and $c_3$ (only 3 parameters for the case of 4 bins, and for the
cases of 16 to 64 bins the remaining divisional parameters have been fixed
at 0.5) on a fixed grid and assumed $\ell_{max}=2000$ 
(as well as the noise and beam
characteristics of the Planck 100~GHz channel).
We see that using 64 bins instead of all the 1999 $\ell$-values
in the interval $[2, 2000]$ gives an immense reduction of computing time
at the cost of only a $0.7\%$ increase in the variance.
Even smaller numbers of bins still give a quite acceptable
increase of variance. 

As an explicit example, the maximum value of 
$\cos^2 \theta$ for 64 bins was reached for the following parameter values:
$a_0 = b_0 = c_0 = 0.2$, $b_1 = c_3 = 0.4$, and $c_1 = c_2 = 0.5$.
This corresponds to the following bin separator values for $\ell$:
(2, 3, 5, 7, 9, 11, 13, 15, 17, 25, 33, 41, 49, 57, 65, 73, 81, 101, 121, 
141, 161, 181, 201, 221, 241, 261, 281, 301, 321, 341, 361, 381, 401, 441, 
481, 521, 561, 601, 641, 681, 721, 761, 801, 841, 881, 921, 961, 1001, 1040, 
1088, 1136, 1184, 1232, 1280, 1328, 1376, 1424, 1496, 1568, 1640, 1712, 
1784, 1856, 1928, 2000).

\begin{table} 
\begin{tabular}{ll} 
\hline 
Number bins & Overlap ($\cos ^2\theta $) \\ 
\hline 
4 & 0.608 \\ 
8 & 0.846 \\ 
16 & 0.925 \\ 
32 & 0.974 \\ 
64 & 0.993 \\ 
\hline 
\end{tabular} 
\caption{Quality of estimator as a function of number of bins
for $\ell_{max}=2000$.} 
\label{TableThree} 
\end{table}

\subsection{Validation of estimator}
\label{validation}

To test our binned estimator, we applied it to 100 CMB-only maps with
$\ell_{max}=2000$ and $n_\mathrm{side}=1024$, kindly
provided by Michele Liguori based 
on the WMAP5-only best-fit parameters \citep{liguori}.
To these maps we added the effects of uniform white noise and a Gaussian 
beam according to the specifications of the 100, 143, and 217~GHz
channels of the Planck experiment (combined in quadrature) \citep{bluebook}.
We have used 64 bins, with divisional 
parameters such that $\cos^2 \theta$ is maximal.

\begin{figure}
\includegraphics[width=9cm]{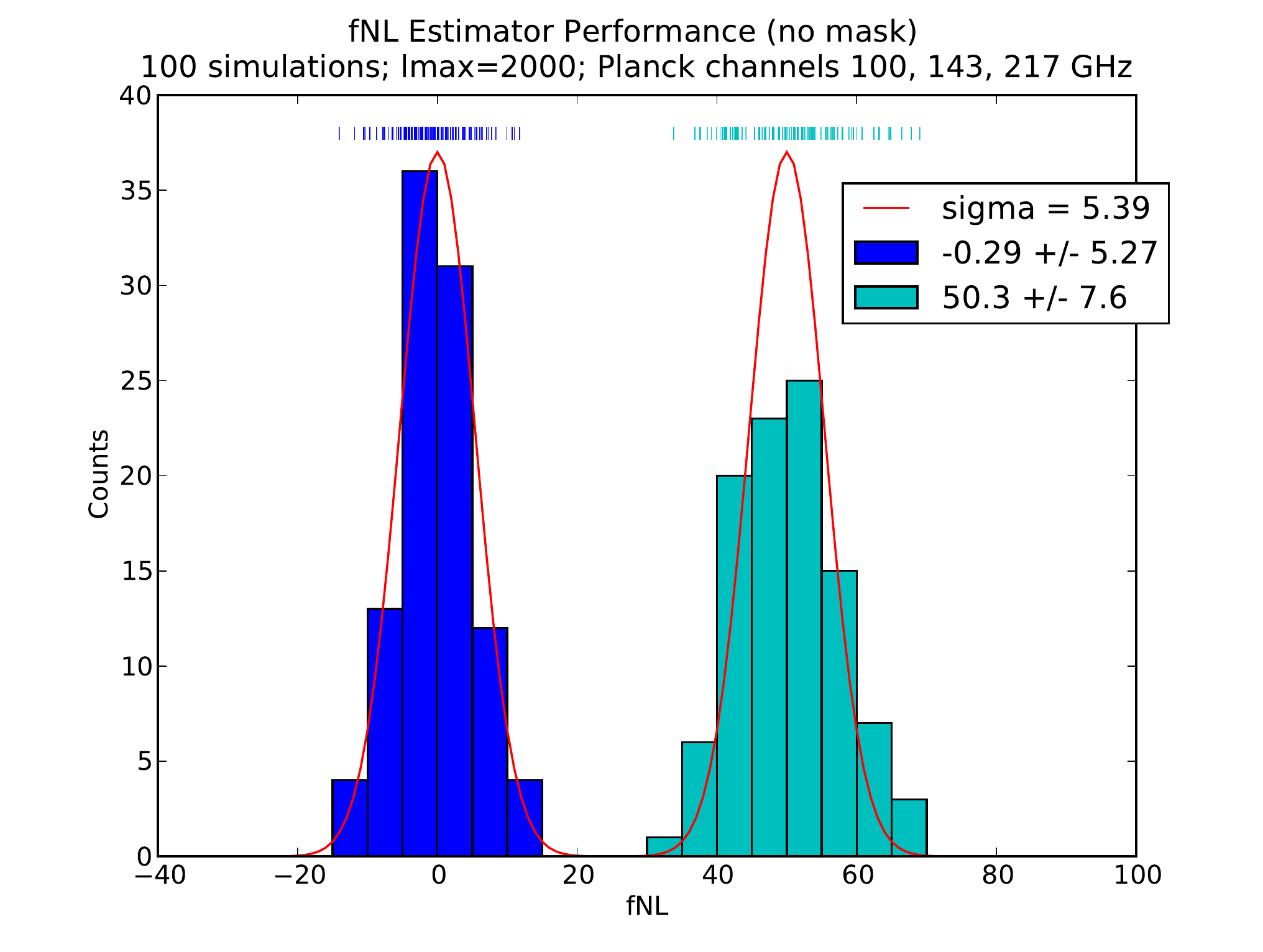}
\includegraphics[width=9cm]{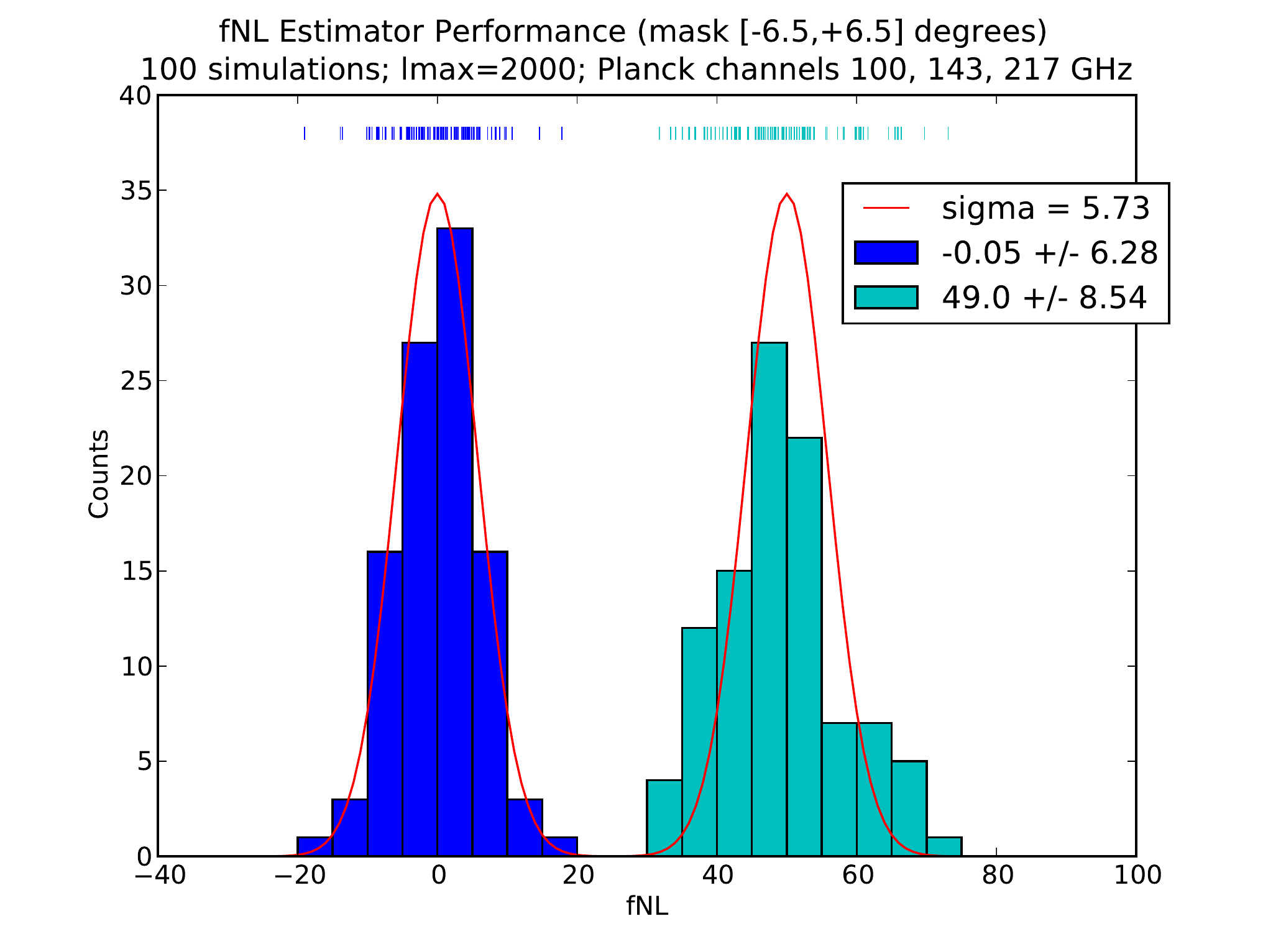}
\caption{
The performance of the binned estimator described
in the text is demonstrated on two sets
of mock maps generated by Monte Carlo, one
with $f_{NL}=0$ and another with $f_{NL}=50.$
In the upper panel the entire sky is used
whereas in the lower panel a mask of $\pm 6.5^\circ $
about the galactic equator has been used, resulting
in a modest increase in estimator variance. 
Gaussian fits to the theoretical mean and variance are
indicated in the histograms.
}
\label{fNL_est_performance}
\end{figure}

The results can be found in Fig.~\ref{fNL_est_performance}.
Without a mask, we recover for $f_{NL}=0$ a value of $-0.3 \pm 5.3$ and for 
$f_{NL}=50$ of $50.3 \pm 7.6$. 
These results agree with those obtained with the KSW estimator.
We also tested with a simple band mask of $[-6.5, +6.5]$ 
degrees around the galactic equator 
and found $-0.05 \pm 6.3$ and $49.0 \pm 8.5$, respectively.
Here the masked area was filled with the average of the rest 
of the map and any remaining monopole was subtracted. A more detailed
study of the influence of masks and how to treat them will be the subject 
of a future publication.

The smooth curve in both plots is the predicted Gaussian distribution
based on the calculated theoretical variance when nonlinear corrections
are ignored. Its standard deviation is $5.4$ without a mask and $5.7$ with
the band mask. The modest increase with the mask is due to the decreased sky 
fraction.
We see that for $f_{NL}=50$, which corresponds to a detection
of several sigma, the actual standard deviation starts deviating from
the Gaussian one as nonlinear corrections become important.

\section{Discussion}

The search for primordial non-Gaussianity is presently in its early stages.
Non-Gaussianity of the local type is motivated by multiple-field inflation 
and if detected would have a substantial impact on our understanding
of the physics of the early universe. At present we have only
an upper limit with $f_{NL}^{local}$ in the interval $[-9,111]$ at
95\% confidence according to the official WMAP5 analysis \citep{wmap5},
which constitutes a hint of a detection. A detection has been claimed
in another analysis of the WMAP data by \cite{yadav},
and several other alternative analyses, such as 
using needlets \citep{needlets,needlets_b} or
spherical Mexican hat wavelets \citep{mexican_hat},
have been published. 
See also \cite{hikage,wmapCrem,wmapCremBis,current_limits,smith}.
These analyses differ in detail but are broadly 
consistent. Note that a confirmed detection of a primordial $f_{NL}$ 
of order 10 would rule out the simplest single-field slow-roll inflation 
models.

The PLANCK space mission promises to bring about approximately
an order of magnitude improvement in sensitivity to $f_{NL},$
so if the present hint of a signal is confirmed, we will have 
an abundance of signal-to-noise at our disposal, allowing the data 
to be divided in many ways in order to confirm the predicted
shape for the bispectrum and to exclude other non-primordial 
sources of parasite bispectral power. 

In this paper, we demonstrated how a binned bispectrum
estimator can be useful to evaluate the perfect template
with a modest number of bins, so that applying the estimator
is computationally extremely efficient. Compared to 
other implementations of the optimal estimator, which are
likewise fast, the present estimator has the advantage that
it can be used to apply other templates, either for $f_{NL}$ or to
determine other bispectral characteristics.
It will be important to develop models of the 
expected pattern of bispectral non-Gaussianity arising
from foregrounds, instrumental artifacts, and foreground
subtraction residuals. Even if very approximate, such
templates will be invaluable for excluding non-primordial
explanations for a possible future detection. We showed
how using a binned estimator one should modify 
the template for primordial non-Gaussianity of the local type to 
optimally mask the contaminants.  

We also investigated how the acoustic oscillations manifest
themselves in the CMB bispectrum from local primordial non-Gaussianity
and in particular evaluated the signal-to-noise for detecting
various features.
Although a rich pattern of non-trivial
acoustic oscillations occurs in the central region of triangle
shape space, we found that these oscillations contribute
negligibly to the total available signal-to-noise. However,
the oscillations having the same shape as the power spectrum
near the vertices of triangle shape space offer a much larger
signal. We also investigated broader features due to the damping,
and these may be seen with PLANCK if $f_{NL}$ is large enough.
An observation of such features in the bispectrum matching
the shape of the two-point power spectrum would constitute
a convincing confirmation of a first $f_{NL}$ detection.
The absence of a signal in other regions where none is 
expected also provides a powerful check against foreground
and instrumental contamination. 

\section*{Acknowledgments}

We would like to thank Michele Liguori for kindly providing us with simulated
sky maps with non-zero $f_{NL}$ that were invaluable for testing
and validating our estimator. We would also like to thank 
Michele Liguori for useful discussions. We thank 
David Spergel for some insightful comments on the final manuscript. 

{\em Note added:} 
After this manuscript was submitted to astro-ph, a paper by 
Fergusson, Liguori and Shellard 
\citep{JamesMichelePaul} appeared
on $f_{NL}$ estimation
having some overlap with
the ideas developed in this paper.

\end{document}